\documentclass[aps,pre,twocolumn,superscriptaddress,amsmath,amssymb,longbibliography,floatfix,nofootinbib]{revtex4}

\usepackage{graphicx}

\usepackage{enumitem,kantlipsum}

\usepackage{bm}
\usepackage[normalem]{ulem}
\usepackage{color}
\usepackage{gensymb}

\usepackage{epstopdf}

\usepackage{xcolor}
\usepackage{color}

\definecolor{green}{rgb}{0,0.5,0}

\begin{document}
 
\title{Nonequilibrium steady states in coupled asymmetric and symmetric exclusion processes}
\author{Atri Goswami}\email{goswami.atri@gmail.com}
\affiliation{Gurudas College, 1/1, Suren Sarkar Road, Jewish Graveyard, Phool Bagan, Narkeldanga, Kolkata 700054, West Bengal, India }
\author{Utsa Dey}\email{utsadey41@gmail.com}
\affiliation{Barasat Government College,
10, KNC Road, Gupta Colony, Barasat, Kolkata 700124,
West Bengal, India}
\author{Sudip Mukherjee}\email{sudip.bat@gmail.com, aca.sudip@gmail.com}
\affiliation{Barasat Government College,
10, KNC Road, Gupta Colony, Barasat, Kolkata 700124,
West Bengal, India}

\begin{abstract}
We propose and study a one-dimensional (1D) model consisting of two lanes with open boundaries. One of the lanes executes diffusive
and the other lane driven unidirectional or asymmetric exclusion dynamics, which are mutually coupled through particle exchanges in the bulk. We elucidate the generic nonuniform steady states in this model. We show that  in a parameter regime, where hopping along the TASEP lane, diffusion along the SEP lane and the exchange of particles between the TASEP and SEP lanes compete, the SEP diffusivity $D$ appears as a tuning parameter for both the SEP and TASEP densities  for a given exchange rate in the nonequilibrium steady states of this model. Indeed, $D$ can be tuned to achieve phase coexistence in the asymmetric exclusion dynamics together with spatially smoothly varying density in the diffusive dynamics in the steady state. We obtain phase diagrams of the model by using mean field
theories,  and corroborate and complement the results by stochastic Monte Carlo simulations. This model reduces to an isolated open totally asymmetric exclusion process (TASEP) and an open TASEP with bulk particle nonconserving Langmuir kinetics (LK), respectively, in the limits of vanishing and diverging particle diffusivity in the lane executing diffusive dynamics. Thus this model works as an overarching general model, connecting  both pure TASEPs and TASEPs with LK in different asymptotic limits.  
We further define phases in the SEP and obtain phase diagrams, and show their correspondence with the TASEP phases.
In addition to its significance as a 1D driven, diffusive model, this model also serves as a simple reduced model for cell biological transport
by molecular motors undergoing diffusive and directed motion inside eukaryotic cells.
\end{abstract}

\maketitle

\section{Introduction}

Natural systems driven by an external field or containing collections of self-propelled particles form prominent  examples of nonequilibrium systems that often
evolve into stationary states carrying steady currents. 
The presence of steady currents distinguish these systems  from their counterparts in thermal equilibrium.
Understanding the general physical principles behind such nonequilibrium transport has been the subject of
intense research recently. One is often particularly interested in nonequilibrium transport
in the context of simple one-dimensional (1D) model systems. In order to elucidate the nature of such nonequilibrium
steady states and in the absence of a general theoretical framework, it is useful to study purpose-built simple models. To this end, a variety of driven lattice gas models have been
introduced and studied extensively~\cite{basic}.

The Totally Asymmetric Simple Exclusion Process (TASEP) with open boundaries is one of the simplest 1D nonequilibrium models, which displays
 boundary induced phase transitions. It was originally
proposed by MacDonald {\em et al} \cite{mac}  to model
the transport of ribosomes along messenger RNA strands in cell biological context. Subsequently, it was reinvented as a paradigmatic nonequilibrium model~\cite{krug}. TASEP consists of a one-dimensional (1D) lattice, along which particles can stochastically hop
from left to right, at rate unity provided the target site is vacant.
The interaction between the particles is through a hard core exclusion potential. The latter ensures
only single occupancy per lattice site, which implies exclusion. In TASEP, particles can enter the system from the left boundary
and exit through the right with certain prescribed entry ($\alpha$) and exit ($\beta$) rates. The phases of TASEP can be tuned by
$\alpha \leq 1$ and $\beta \leq 1$, and are generically uniform in space, except for a special case when
the two rates are equal and less than 1/2. Three distinct phases can be identified in the
phase diagram of TASEP constructed in the space spanned by the control parameters $\alpha$ and $\beta$. 
These are the High density (HD) phase, Low density (LD) phase and the Maximal current (MC) phase.
 TASEP is one of the very few nonequilibrium models which
can be exactly solved and has emerged as a simple basis to study the principles and phenomenologies of 1D transport~\cite{privman,tasep-rev,phases,tasep}.

The Symmetric Exclusion Process (SEP) is a simple realisation
of the 1D {\em equilibrium diffusion process} in which particles can move,
in contrast to TASEP, in either direction (left or right) symmetrically, subject to exclusion. Also unlike TASEP,
the entry and exit of particles can occur at both the ends of the lattice. In the steady state,
the spatial dependence of the density profile is always a straight line, either fully flat for
equal biases or an inclined line in case of unequal biases \cite{sep}.

More recently,  TASEP has been generalised in a variety of ways, all of which 
reveals many new interesting macroscopic phenomena. These usually involve the presence of additional  microscopic processes competing with the basic hopping process of TASEP, or presence of conservation laws. A prominent example is a model introduced in Ref.~\cite{frey-lktasep}, which has competing 1D nonequilibrium transport (TASEP)
and equilibrium on-off exchanges with a surrounding reservoir (known as Langmuir kinetics (LK)).
   In LK, one studies the attachment-detachment
kinetics of particles on a lattice coupled to a bulk reservoir. As a physical motivation, this provides the simplest description of binding and unbinding kinetics of enzymes to some substrate. In LK dynamics, the particles get adsorbed
at an empty site or detached from an occupied one with some given rates. As in TASEP,
the only interaction between the particles is the hard-core repulsion due to particle exclusion, leading to maximum occupancy one per site even in the presence of LK.
The LK and TASEP are two of the simplest paradigmatic equilibrium and nonequilibrium models, which clearly contrast
equilibrium and non-equilibrium dynamics, distinguishing the corresponding stationary states. For instance, LK maintains detailed balance, and evolves
into an equilibrium steady state in the long time limit. In contrast, a TASEP naturally breaks the detailed balance condition due to continuous flow of particles, and the resulting  stationary state is a
non-equilibrium state that carries a finite current. Such non-equilibrium steady
states are known to be quite sensitive to changes in the boundary conditions. In contrast, equilibrium steady
states are very robust to such changes and dominated by the bulk dynamics. TASEP is a boundary condition dependent process - in the TASEP
new particles can enter or leave the system only at the system boundaries, whereas
in the bulk there are no sources or sinks. In contrast in LK particles
can enter or leave the system at any site. As shown in Ref.~\cite{frey-lktasep}, a combination of the two can produce nonuniform steady state
densities in the TASEP when the typical    {\em exchange
rate} of a particle moving along the TASEP lane is comparable with the entry-exit rates in
the filament, which can be achieved by introducing  system size-dependent LK exchange rates between the bulk and the TASEP lane. When the two rates are comparable, the resulting steady
state density profiles can show coexistence phases and domain walls, in contrast to the 
density profiles in isolated TASEP and Langmuir kinetic processes.

Diffusive processes are ubiquitous in nature, e.g., in cell biological contexts. How diffusive and driven processes may combine to influence mass transport is a fundamentally important question in cell biology. Notable previous studies on this topic include the work on the coupled dynamics of diffusive (unbound) motors
in the cell cytoplasm and motors driven along microtubules (bound motors) in tube-like cylindrical compartments (representing the
cytoplasm), containing one filament along the axis (representing a microtubule)
with motors being injected and extracted at the boundaries~\cite{lipowsky}.
This model
reveals a phase behavior similar to that of 1D TASEP. Later, an extension of the above
model was studied in Ref.~\cite{santen1}. These are however
three dimensional models, which are relatively difficult to analyse, either
analytically or by computer simulations. Moreover, the competition between the time scales of
diffusive and directed dynamics has also not been studied in these works. A 1D {\em closed} model consisting of two equal segments with one segment executing driven dynamics and the other diffusive was studied in Ref.~\cite{hinsch}. Interestingly, unlike an open TASEP, this model shows a single {\em localised domain wall} (LDW) instead of a delocalised domain wall (DDW) found in an open TASEP. This is attributed to the overall particle number conservation in the model. Very recently, it was shown that in the limit of a large diffusive segment, an LDW in this model can get delocalised due to fluctuation effects~\cite{parna}.

Our motivation here is to systematically investigate the interplay between the diffusive, driven and particle-exchange time-scales in 1D,  subject to exclusion. We do this by generalising 1D nonequilibrium transport by coupling
TASEP with SEP via particle exchange that is reminiscent of LK. We also discuss effects of space-dependent exchanges on the steady states.
We expect the steady states of a coupled SEP-TASEP model will be quite different from the features of the decoupled systems, i.e., of an isolated TASEP and an isolated SEP.
As we shall see, for our coupled system in the steady state we find phase co-existences in 
TASEP and spatially non-uniform (but smooth) density profiles in SEP, depending upon relative time 
scales of SEP and TASEP dynamics. This is totally in contrast to the well-known
spatially uniform densities in steady states of isolated  TASEP
and SEP. {The effects of combining driven and diffusive transport have been studied earlier in various contexts. For instance, Ref.~\cite{hinsch} considered a closed system consisting of a TASEP and a SEP segments, which showed localised domain walls in the TASEP for some choices of the model parameters. Subsequently, this was studied in a half-open geometry~\cite{graf1,graf2}. See also Ref.~\cite{evans} for a related study on a two-lane system coupling an asymmetric and symmetric exclusion processes.}
Nonetheless, a systematic understanding of the effects of the competition between different time scales is clearly lacking. Lastly, space dependence of the parameters which define the local dynamics are expected to play an important role in controlling the nature of the steady states of driven systems, see, e.g., Refs.~\cite{tirtha-qkpz,sudip-jstat} and references therein for detailed studies on the steady states in periodic and  open TASEPs with smoothly space-dependent hopping rates.   Indeed, spatially smoothly varying rates of particle exchanges between the TASEP and SEP lanes naturally extend the studies reported in Refs.~\cite{tirtha-qkpz,sudip-jstat}. 

{ The principal results in this work are as follows:

(i) For a finite diffusivity in the SEP channel and equal attachment-detachment rates between the TASEP and SEP channels, both the TASEP and SEP steady state density profiles acquire complex space dependence. At the location of a discontinuity in the TASEP density, the SEP density has a strong space dependence.

(ii) For a diverging SEP diffusivity, the TASEP density profiles are identical to those in the well-known model of an open TASEP with LK. In the same limit, the SEP density profiles become flat with a value of $1/2$.

(iii) For a vanishing SEP diffusivity, the TASEP density profiles reduce to those of an open isolated TASEP, whereas the SEP density profiles in the bulk strictly follow the TASEP density profiles.

(iv) The TASEP and SEP phase diagrams are shown to have a strong correspondence with each other.

(v) As the SEP diffusivity is reduced, a domain wall in the TASEP channel gets gradually delocalised.

 At a broader, overarching level, the novelty of our coupled SEP-TASEP model lies in the fact that not only it generalises the TASEP-LK and pure TASEP models, it also shows by changing the SEP diffusivity, {\em a SEP parameter}, one can vary the TASEP density, and induce a diffusion-controlled delocalisation transition of a localised domain wall in the TASEP channel.} 

Apart from its  significance in nonequilibrium statistical mechanics,
our model in addition has a biological inspiration as well: it may be viewed
as a simple model for interacting bound and unbound molecular motors
inside eukaryotic cells. Molecular motors inside eukaryotic cells transport cargo and are responsible for almost 
all kinds of cellular motility \cite{motor}.
These are inherently nonequilibrium processes, sustained by 
the energy released in the hydrolysis of
ATP (Adenosine triphosphate), producing ADP (Adenosine 
diphosphate) and inorganic phosphate.
The molecular motors typically hop  unidirectionally
along the microtubules in cells. Examples of
 such molecular motors are the processive motors belonging to the kinesin 
family. However, due to fluctuations of both thermal and non-thermal
origin,  in course of their unidirectional 
hopping motion, these molecular motors can stochastically detach off to the surrounding cytoplasm.
In the cytoplasm, these molecular motors  
diffuse around until they again get themselves attached to a filament.
The cytoplasm thus effectively acts as a reservoir for these (unbound or detached) 
molecular motors. On the whole, thus, the bound molecular 
motors hop unidirectionally along the filaments, whereas the unbound 
molecular motors in the cytoplasm undergo diffusive motion, and these two
can stochastically exchange molecular motors between them. We here construct and study a simple one dimensional model that reproduces these collective 
behaviours of transport in a cell.


The rest of this article is organised as follows. In Sec.~\ref{model}, we introduce our model. Next, in Sec.~\ref{mft} we set up the mean field theory for our model. Then in Sec.~\ref{mcs} we briefly discuss the stochastic simulations performed. Next, in Sec.~\ref{steady} we extensively present and analyse the steady state densities and the associated phase diagrams in both the TASEP and SEP channels. In this Section, we  study the nonequilibrium steady states with constant attachment detachment rates, together with a rate of symmetric hopping or diffusion that remains finite relative to the attachment-detachment rates of the LK dynamics, or the unidirectional hopping rates along the TASEP lane. Results from both mean field theory (MFT) and Monte Carlo simulation (MCS) studies are presented. In Sec.~\ref{ddw}, we illustrate gradual delocalisation of the domain walls as the diffusivity is reduced. In Sec.~\ref{summ}, we summarise and discuss our results. In Appendix, we present  our mean-field results in another limit of the model, {\em viz.}, space-dependent attachment-detachment rates together with a diffusive time scale that diverges relative to the typical particle residence time due to the LK dynamics.

\section{Model}\label{model}





\begin{figure}[htbp]
\centering
\includegraphics[width=0.5\textwidth]{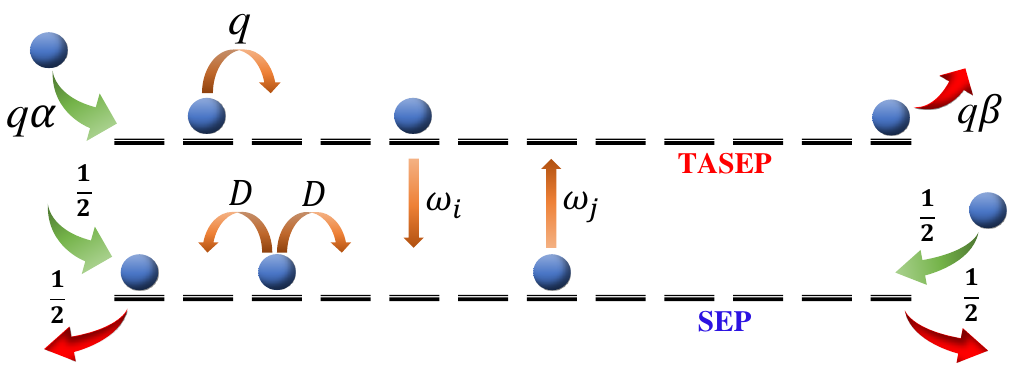}
\caption{ Illustration of the two-lane model. We label the
upper lane as the TASEP lane and the lower one as the SEP lane.
Particles on the upper lane follow  TASEP dynamics with {hopping rate $q$} in the bulk subject to exclusion and entry and exit rates 
{$q\alpha$ and $q\beta$}, respectively. Particles on lower lane  obey SEP dynamics  
{with hopping rate $D$}
and possess entry rates $1/2$,  and also exit rates $1/2$, at
the left and right end, respectively. The local space-dependent exchange rate between the
lanes is denoted by $\omega_i$ that can depend on the site $i$.}
\label{fig:Model}
\end{figure}

In this work, we investigate the nature of the non-equilibrium
steady states of a  coupled two-lane model consisting of two equally sized lanes with $L$ 
lattice sites each, whose dynamics is governed by  
a totally asymmetric exclusion and a symmetric exclusion processes, respectively; see Fig.~\ref{fig:Model}. The sites in each lane are labelled by the index $i=1,..,L$. The dynamical update rules of this model consist of the following steps.

(a) {Particles on the top lane  (TASEP) may enter at the left end ($i=1$) at rate $q\alpha$, stochastically hop from $i=1$ to $L$ unidirectionally subject to exclusion at rate $q$, and leave
the lane at the right end ($i=L$) at rate $q\beta$~\cite{sudip-jstat}. (Note that in a conventional study on pure open TASEP, usually $q$ is set to unity.) }

(b) On the bottom lane  (SEP) particles
hop with equal rate $D$ in either direction subject to exclusion, and also 
may leave or enter this lane at the right ($i=L$) and left ($i=1$) end at rate 1/2.  

(c) Particles hopping on these parallel tracks may detach
from one lane and attach to the other, with generally site-dependent but equal attachment and detachment rates $\omega_i$.

All these hopping and exchange processes in both the lanes are 
allowed under the strict constraint of an \textit{on-site exclusion principle},
which forbids double-occupancy of particles in any of the lattice sites.

\paragraph{Time scales:}  The different time scales for our model are mentioned below :\\
$\bullet$ $\tau_\text{TASEP} = {q}$ : Time-scale of the directed dynamics of the particles on the TASEP lane. 
This sets the time-scale for our model.
\\
$\bullet$ $\tau_\text{SEP} = D$ : Time-scale of the diffusive dynamics of the 
particles on the SEP lane. 
\\
$\bullet$ $\tau^\times_{\,i} = \omega_i^{-1}$ : Time-scale of the lane exchange mechanism which 
couples the filament to the surrounding reservoir. 
\paragraph{Symmetries:}  This model admits   the {\em particle-hole symmetry}, which
will prove helpful in constructing and understanding the phase diagrams for the filament and the reservoir lanes.
We note that the absence of a particle at any site on the two lanes can be interpreted as the presence of a 
vacancy or a hole at that position. A particle hopping from the left site to the empty lattice
site to its right in the bulk may be considered as a hole hopping form the right to the left
lattice site. Likewise, the entry of a particle from the left end of the lattice can be 
considered as an exit of a hole and vice-versa. Similarly for the particle exchange dynamics between the TASEP and SEP lanes, movement of a particle from (to) TASEP to (from) SEP may be viewed as a hole moving to (from) TASEP from (to) SEP.
In fact, formally the model remains invariant under the
transformation of all particles into holes, with a discrete transformation of sites 
$i \leftrightarrow L - i$ and all pairs of the entry-exit rate, e.g. $\alpha \leftrightarrow \beta$. These define the {\em particle-hole symmetry} in this model.
As a consequence of this  symmetry, the phase diagram in the $\alpha-\beta$ plane
can be split into two complementary regions by the $\alpha = \beta$ line. As a result, it suffices
to understand the phase diagram for only one of the two regions.
The phase behaviour of the system in the remaining region can be constructed and analysed
by using the particle-hole symmetry.

\section{Mean-field theory}\label{mft}

The microscopic dynamics of TASEP is prescribed by the rate equations for
every site in the SEP and TASEP lanes. The MFT approximation entails neglecting the correlation effects and replacing the average of product of the densities by the product of average of densities in the steady states~\cite{tasep-rev}. Although this is an uncontrolled approximation, this has been found to work with high degree of accuracy in the original TASEP problem and its many variants subsequently (see, e.g., Refs.~\cite{frey-lktasep,niladri1,tirtha1} as representative examples); we use the MFT here as a guideline in our analysis below.

The dynamical equations of motion for $\rho_i$ and $c_i$, the TASEP and SEP densities at site $i$ in the TASEP and SEP lane respectively, are
\begin{eqnarray}
 \partial_t \rho_i &=& {q}\rho_{i-1}(1-\rho_i) - {q} \rho_i (1-\rho_{i+1}) \nonumber \\&+& \omega_i [ c_i(1-\rho_i)-\rho_i (1-c_i)],\label{beq1}\\
 \partial_t c_i&=& D[(c_{i-1}+c_{i+1})(1-c_i) - c_i(2-c_{i-1}-c_{i+1})] \nonumber \\&-& \omega_i [ c_i(1-\rho_i)-\rho_i (1-c_i)]. \label{beq2}
 \end{eqnarray}
It is easy to see that the MFT equations (\ref{beq1}) and (\ref{beq2}) are invariant under the particle-hole symmetry defined above. 
 
To proceed further, we first take the continuum approximation: we take $L\rightarrow \infty$, which makes $x$ a continuous variable between 0 and 1: $x\in [0,1]$. Without any loss of generality, we assume unit geometric length for the whole lattice (both TASEP and SEP), and define a lattice constant $\varepsilon = 1/L$ that approaches zero as $L\rightarrow \infty$. Thus, in the thermodynamic limit, $\varepsilon$ serves as a small parameter.  Further, we define $\rho(x)=\langle \rho_i\rangle$ and $c(x)=\langle c_i\rangle$ as the steady state densities of TASEP and SEP respectively at $x$. In the steady state, we expand the different terms on rhs of (\ref{beq1}) and (\ref{beq2}) in a Taylor series in powers of $\varepsilon$ to obtain
 \begin{eqnarray}
  0&=& \omega (x)(c-\rho)+ \frac{{q}}{L}(2\rho-1)\partial_x\rho + \frac{{q}}{2L^2}\partial_x^2 \rho,\label{rho-eq}\\
  0&=& \omega (x) (\rho-c)+ \frac{D}{L^2}\partial_x^2 c.\label{c-eq}
 \end{eqnarray}
Here, we have retained terms up to ${\cal O}(\varepsilon^2)\equiv {\cal O}(1/L^2)$ in the Taylor expansions above, discarding all higher order terms. We note the different $L$-dependence of the terms in (\ref{rho-eq}) and (\ref{c-eq}). {In order to make the nonconserving Langmuir kinetics terms compete with the hopping terms in (\ref{rho-eq}) and diffusion in (\ref{c-eq}), we define $\omega\equiv \Omega/L^2$, where $\Omega\sim {\cal O}(1)$~\cite{frey-lktasep}, and set $q=1/L$. Thus with $q=1/L$, particles enter and exit the TASEP channel at effective rates $\alpha/L$ and $\beta/L$, and hop along the TASEP channel at rate $1/L$.} With these parameter rescalings, we obtain the steady state in the thermodynamic limit $L\rightarrow\infty$
\begin{eqnarray}
 &&\Omega (x)(c-\rho)+(2\rho-1)\partial_x\rho =0,\label{rho-mft}\\
 && D\partial_x^2 c + \Omega (x) (\rho-c)=0. \label{c-mft}
\end{eqnarray}
Equations (\ref{rho-mft}) and (\ref{c-mft}) are the starting point of the MFT for this model, which we discuss next.


We note that the full MFT equations (\ref{rho-mft}) and (\ref{c-mft}) have the following well-known limits. 
{Indeed, there are two limits in which the MFT equations (\ref{rho-mft}) and 
(\ref{c-mft}) reduce to two well-known models whose MFT solutions are already known.}  These two limits are characterised by the limiting values of  $\Omega/D$.
 Consider now $\Omega(x)$ as constant, $\Omega(x)=\Omega$ at all $x$ together with

(i) $D\rightarrow \infty$, i.e., $\Omega/D$, for a given $\Omega\sim {\cal O}(1)$, vanishes. In this limit from (\ref{c-mft}), assuming $c(0)=1/2=c(1)$, $c(x)=1/2$ everywhere. Substituting this in (\ref{rho-mft}), we get
\begin{equation}
    \frac{\Omega}{2}(1-2\rho)+(2\rho-1)\partial_x\rho =0.\label{rho-mft-1}
\end{equation}
This is identical to the MFT equation for $\rho(x)$ in the LK-TASEP problem with an equal attachment-detachment rate of value $\Omega/2$~\cite{frey-2lane}. Physically, as $D$ diverges, the diffusive dynamics in the SEP lane becomes extremely fast, effectively reducing 
attachment-detachment events to the SEP lane insignificant relative to the in-lane diffusion, over TASEP hopping and attachment-detachment time-scales. This means the average steady state density in the SEP lane is 1/2, independent of the precise values of the attachment-detachment rates, as we found above. This in turn means that the attachment-detachment events at rate $\Omega$ to (from) the TASEP will take places with a background SEP density of 1/2, unaffected by the TASEP dynamics, as we see form (\ref{rho-mft-1}).

(ii) $D\rightarrow 0$, i.e., $\Omega/D$ diverges for a given $\Omega\sim {\cal O}(1)$.  In this limit from (\ref{c-mft}), $c(x)=\rho(x)$ is the solution in the bulk. Substituting this in (\ref{rho-mft}), we get
\begin{equation}
 (2\rho-1)\partial_x\rho =0,
\end{equation}
which is nothing but the MFT equation for the steady state density in the standard TASEP problem, giving the LD, HD and MC phases~\cite{tasep-rev}. Actually for vanishingly small $D$, the only dynamics in SEP are the attachment-detachment events, which have the effects of locally decreasing the differences in the densities in the TASEP and SEP lanes. Indeed, when $D\rightarrow 0$, different sites of the SEP lane virtually decouple from each other, and only exchange particles with the corresponding site in the TASEP lane having a density $\rho(x)$. In the steady state then $c(x)=\rho(x)$ is achieved, no further time evolution of the SEP density. We can argue this in yet another way. As mentioned above, in the limit $D\rightarrow 0$, different SEP sites are effectively decoupled from each other; the only dynamics operating is the exchange with the corresponding TASEP sites at equal attachment and detachment rates. For a SEP site at $x$ in the bulk, that exchanges particles with the corresponding TASEP site at $x$, the steady state density is $c(x)=\rho(x)$. When this is achieved, there is no more net flow of particles between the SEP and TASEP channels {\em on average}, as there cannot be any net particle exchanges on average between two sites having the same average densities. Thus, on average the exchange term becomes irrelevant in the TASEP mean-field equation in the steady states, giving effectively just the pure TASEP mean-field equation.~\footnote{It should be noted that instantaneously (i.e., not on average) the SEP and TASEP densities at any given SEP or TASEP site are in general unequal and particle exchange continues to operate at any instant between the SEP and TASEP sites.}

We thus find that for a fixed $\Omega$, a key parameter which controls the shape of the density profiles on both the TASEP and SEP
lanes is the magnitude of the effective diffusion constant $D$. If diffusion
of the SEP lane is very slow, $D \to 0$, we find from Eq.~(\ref{c-mft}) that 
the density on that reservoir lane becomes effectively slaved to the density on the 
filament, $c(x)=\rho(x)$. Hence, in this limit, the filament dynamics is
independent of the reservoir and simply given by that of the TASEP. 
In contrast, in the opposite limit where $D$ is large, i.e., diffusion on
the reservoir lane is fast, $c(x)$ becomes independent of $\rho(x)$ and
shows a flat profile, e.g. $c(x)=1/2$ with $c(x=0)=c(x=1)=1/2$, as the boundary conditions. In this case the
SEP lane simply acts as a reservoir with a constant particle density
similar to the TASEP-LK model with an attachment rate $\Omega_A = \Omega/2$
and a detachment rate $\Omega_D = \Omega/2$, respectively \cite{frey-lktasep}. Notice that independent of $\Omega(x)$ and $D$, $\rho(x)=1/2=c(x)$ remain solutions of the MF equations (\ref{rho-eq}) and (\ref{c-eq}).

As an aside, we also note that solving the full MFT equations (\ref{rho-mft}) and (\ref{c-mft}) with space-dependent $\Omega(x)$ and an arbitrary $D$ is analytically challenging and also not particularly insightful. Instead, we solve (\ref{rho-mft}) and (\ref{c-mft}) in the following cases: (i) $\Omega/D$ finite but small with $\Omega$ being independent of $x$, (ii) $\Omega/D$ finite but large with $\Omega$ being independent of $x$. We also briefly consider  $\Omega(x)$ to be space varying, but assume $D$ diverges, $\Omega/D$ vanishes at all $x$. The MFT equation for $\rho(x)$ now becomes
\begin{equation}
    \Omega(x)(\frac{1}{2}-\rho)+(2\rho-1)\partial_x\rho =0.\label{rho-mft-2}
\end{equation}
This is the MFT equation for the LK-TASEP problem but with an equal, space varying attachment-detachment rate. This is discussed in Appendix.

\section{MONTE CARLO simulation}\label{mcs}

The TASEP and SEP lanes of the model consist of $L$ sites each, labelled by an index $i$ with $i\in [1,L]$.
 Let $\rho_i(t)$, which is either 0 or 1, be the occupation at site $i$ of the TASEP channel, and $c_i(t)$, which is again either 0 or 1, be the occupation at site $i$ of the SEP channel  at time $t$. {We perform MCS studies of the model subject to the update rules (a)-(c) described above in Sec.~\ref{model} by using a random sequential updating scheme. The particles enter the system through the left most site ($i=1$) in the TASEP channel at a fixed rate $q\alpha$, subject to exclusion, i.e., if $\rho_1=0$. After hopping through the system from $i=1$ to $L$ at rate $q$, subject to exclusion, the particles exit the system from $i=L$ at a fixed rate $q\beta$. Here, $\alpha$ and $\beta$ are the two simulation parameters, which are varied to produce different steady states. We have chosen $q=1/L$ in our MCS studies. }In the SEP channel, particles can enter at rate $1/2$ if the site $i=1$ or $i=L$ of the SEP channel is vacant, or if it is filled, a particle either leaves the system through the left or right end respectively at rate $1/2$, or hops to the site $i=2$ or $i=L-1$ at rate $D$, if the target site is empty. In general, in the bulk of the SEP channel, a particle can hop to its left or right site with equal probability at rate $D$, provided the target site is empty. 
 {We use $D\leq 1$}. Lastly, we allow exchange of particles between the SEP and TASEP channels at any site $i$, subject to exclusion, at rate $\omega$. After reaching the steady states, the density profiles are calculated and temporal averages are performed.  This produces time-averaged, space-dependent density profiles, given by $\langle \rho_i(t)\rangle$, and $\langle c_i(t)\rangle$; here $\langle...\rangle$ implies temporal averages over steady states. The simulations have been performed with  $L=1000$ up to $10^9$ Monte-Carlo steps. 
 { Lastly, all the measurements are made in the steady states, which are reached by the system after spending certain transient times. 
 
\section{Steady state densities}\label{steady}






In the previous Section, we have discussed  that for a fixed $\Omega$ the diffusion constant $D$ determines the steady state density profiles in both the TASEP and SEP lanes. For intermediate values of  $D$ the density
profiles on both lanes deviate from the known asymptotic results. With increasing the magnitude of $D$ one can study 
the crossover from TASEP to TASEP-LK dynamics, and it will be interesting to 
see how the density profiles and the ensuing phase diagram change as $D$ is varied. Before we proceed to solve the MFT equations, we note the following general result. The microscopic dynamical rules set up in Sec.~\ref{model} above clearly maintain overall particle conservation (i.e., in the combined TASEP and SEP) locally in the bulk of the system, since particle entry or exit events (considering the overall system) take place only at the boundaries, although individually the TASEP and SEP lanes do not conserve particles in the bulk locally due to the particle exchanges between them. This fact is clearly reflected in the MFT equations (\ref{beq1}) and (\ref{beq2}), or in their continuum analogues (\ref{rho-eq}) and (\ref{c-eq}), which can be combined to produce a conservation law. Indeed, the MFT equations (\ref{rho-mft}) and (\ref{c-mft}) can be combined to produce a conservation law given by
\begin{equation}
 (2\rho(x)-1)^2 + 4D \partial_x c = J=1-4j_\text{tot}\label{mft-con}
\end{equation}
where $j_\text{tot}$ is the total current through the SEP and TASEP channels combined (see also later). Equation~(\ref{mft-con}) reveals the following: Since the steady state TASEP density $\rho(x)$ can have at most a finite discontinuity (e.g., at the location of a domain wall), so will $\partial_x c$ at the same location to maintain (\ref{mft-con}). Now since $\partial_x c$ can have at most a finite discontinuity, steady state SEP density $c(x)$ must be continuous everywhere (but not necessarily spatially uniform). Nonetheless, at the location of a discontinuity in $\rho(x)$, $c(x)$ should have a strong space dependence, as opposed to where $\rho(x)$ itself is continuous. We shall see below that our actual MFT solutions for $\rho(x)$ and $c(x)$ bear these features.  


We solve the MFT equations (\ref{rho-mft}) and (\ref{c-mft}) perturbatively, assuming $\Omega/D$ to be large or small.
Given the above discussions, interesting features are expected when $\Omega/D$ is finite (can be large or small or just ${\cal O}(1)$). We then expect $c(x)$ to be neither 1/2, nor equal to $\rho(x)$ in bulk. Likewise, $\rho(x)$ is expected to be neither one of the LK-TASEP solutions, nor standard TASEP solutions in the bulk.  

\subsection{MFT for large $D$}

We first consider ``large but finite $D$'', i.e., small but non-zero $\Omega/D$ for a given $\Omega\sim {\cal O}(1)$. In this limit, we solve the MFT equations by perturbatively expanding around the limiting solutions $\rho(x)=\rho_\text{LK}(x)$ and $c(x)=1/2$. For large but finite $D$, we expect small modifications to $\rho(x)=\rho_\text{LK}(x)$ and $c(x)=1/2$. We thus expect phases in the TASEP lane similar to those reported in Ref.~\cite{frey-lktasep} to emerge.  Furthermore, the exchange of particles between the TASEP and SEP lanes should have the physical effects of {\em reducing} locally the difference in the densities in the TASEP and SEP lanes. This means  whenever $\rho(x)>(<) 1/2$, we expect $c(x)>(<)1/2$. This in turn suggests that the steady state density in the SEP lane should be excess (deficit) relative to 1/2, the steady state density of an isolated SEP with equal entry and exit rates. This picture is consistent with the form of the MF equation (\ref{c-mft}) with $\Omega(x)$ being assumed to be a constant. Since $\partial_x^2 c(x)$ that gives the local curvature of $c(x)$ is less than zero for $\rho(x)>c(x)$, expected in the HD phase, $c(x)$ should resemble, loosely speaking, an inverted ``U'', whereas for $\rho(x)<c(x)$, expected in the LD phase,   $c(x)$ should resemble, loosely speaking, an upward ``U''. These considerations suggest that the SEP channel can have an average density more, less or equal to 1/2. We call these {\em excess, deficit} and {\em neutral} phases of SEP. We will see below that these expectations are borne by our MFT solutions.

To proceed with our MFT solutions valid for $\Omega/D\ll 1$, we write
\begin{eqnarray}
 &&\rho(x)=\rho_\text{LK}(x) + \delta\rho(x), \label{del-rho-def}\\ 
 && c(x) = \frac{1}{2} + \delta c(x). \label{del-c-def}
\end{eqnarray}
Here, $\rho_\text{LK}(x)$ is the well-known solution of the LK-TASEP problem and satisfies
\begin{equation}
 (2\rho_\text{LK}-1)(\partial_x \rho_\text{LK}-\frac{\Omega}{2})=0,
\end{equation}
giving
\begin{equation}
 \rho_\text{LK}(x)=\frac{1}{2}\;\;\;\text{or}\;\;\;\rho_\text{LK}(x)= \frac{\Omega}{2}x+\rho_0.\label{rho-lk}
\end{equation}
Here, $\rho_0$ is a constant of integration, which may be evaluated by using the boundary conditions. We set
\begin{eqnarray}
 &&\rho(0)=\alpha=\rho_\text{LK}(0),\label{bc-1}\\
 &&\rho(1)=1-\beta = \rho_\text{LK}(1).\label{bc-2}
\end{eqnarray}
Furthermore, $\delta\rho(x)$ and $\delta c(x)$ are assumed to be ``small'' deviations from $\rho_\text{LK}(x)$ and $c(x)=1/2$. In particular, $\delta c(x)$ satisfies
\begin{equation}
 \partial_x^2 \delta c(x)+ \frac{\Omega}{D}\left[\rho_\text{LK}+\delta\rho - \frac{1}{2}-\delta c\right]=0.
\end{equation}
We know that in the limit $\Omega/D\rightarrow 0$, $c(x)\rightarrow 1/2$ and hence $\delta c(x)\rightarrow 0$. We can thus write
\begin{equation}
 \delta c(x) = f (\frac{\Omega}{D}),
\end{equation}
with $f(0)=0$. This suggests that to the lowest order in $\Omega/D$, $\delta c(x)$ should scale with $\Omega/D$. This further implies that $\delta \rho(x)$, which vanishes as $\delta c\rightarrow 0$, should also scale with $\Omega/D$ to the lowest order in $\Omega/D$. Therefore, to the lowest order in $\Omega/D$, $\delta c(x)$ follows
\begin{equation}
 \partial_x^2 \delta c(x) +\frac{\Omega}{D}[\rho_\text{LK}(x)-\frac{1}{2}]=0, \label{delta-c-eq}
\end{equation}
where $\rho_\text{LK}(x)$ is given by (\ref{rho-lk}). Since $c(x)=1/2$ at $x=0,1$, we must have $\delta c(x)=0$ at $x=0,1$. If we choose $\rho_\text{LK}(x)=1/2$, we get $\delta c(x)=0$ trivially, giving $c(x)=1/2$ in the bulk. This is not surprising, since $\rho(x)=1/2=c(x)$ is a solution in the bulk. Non-trivial solution for $\delta c(x)$ is obtained if we set $\rho_\text{LK}(x)= (\Omega/2)\, x+\rho_0$. Substituting this in (\ref{delta-c-eq}) and integrating with respect to $x$ twice, we obtain
\begin{equation}
  \delta c(x)=  -\frac{\Omega}{D}\bigg[\frac{\Omega x^3}{12} + (\rho_0-\frac{1}{2})\frac{x^2}{2}\bigg] + \overline c_1x + \overline c_2.
\end{equation}
 Constants $\overline c_1,\,\overline c_2$ are the two constants of integration, which may be evaluated by using the boundary conditions. At $x=0$, $c=1/2$ giving $\delta c(0)=0$. This gives $\overline c_2=0$. We further have at $x=1$, $c=1/2$, giving $\delta c(1)=0$. From this condition we obtain
 \begin{equation}
  \overline c_1=\frac{\Omega}{D}\left[\frac{\Omega}{12}+\frac{1}{2}(\rho_0-1/2)\right]
 \end{equation}
giving
\begin{equation}
 \delta c(x)=\frac{\Omega}{D}\left[\frac{\Omega}{12}(x-x^3)+\frac{1}{2}(\rho_0-\frac{1}{2})(x-x^2)\right].
\end{equation}
Notice that $\delta c(x)$ and hence $c(x)$ depend explicitly on the boundary conditions on $\rho(x)$ through $\rho_0$.
The full solution of the steady state SEP density $c(x)$ is given by
\begin{eqnarray}
 &&c(x)=\frac{1}{2} + \delta c(x)\nonumber \\
 &&=\frac{1}{2} +\frac{\Omega}{D}\left[\frac{\Omega}{12}(x-x^3)+\frac{1}{2}(\rho_0-\frac{1}{2})(x-x^2)\right].\label{full-c-large-D}
\end{eqnarray}
Clearly, $c(x)=1/2$ at $x=0,1$.
Since $\rho_0$, being the boundary condition on $\rho(x)$ either at the left or at the right end, depending on whether we are considering LD or HD phases of the TASEP,  for each of $\rho_\text{LD}(x)$ and $\rho_\text{HD}(x)$, the steady state density profiles in the LD and HD phases, respectively, there are distinct solutions of $c(x)$; see below.

We now solve for $\rho(x)$. We start from Eq.~(\ref{rho-mft}), which may be written as 
\begin{equation}
 (2\rho(x)-1)[{\partial}_x\rho(x)-\hat\Omega]=-\Omega c(x) +\frac{\Omega}{2},
\end{equation}
{ where $\hat\Omega \equiv \Omega/2$ plays the role of the scaled attachment-detachment rate in the TASEP-LK problem~\cite{frey-lktasep}}. Now write $\rho(x)=\rho_\text{LK}(x)+\delta\rho(x)$, where $\rho_\text{LK}(x)$ satisfies (\ref{rho-lk}). Then $\delta\rho(x)$ satisfies the following equation.
\begin{equation}
(2\rho_\text{LK}-1){\partial}_x\delta\rho=-\Omega\frac{\Omega}{D} \left[\frac{\Omega}{12}(x-x^3)+\frac{1}{2}(\rho_0-\frac{1}{2})(x-x^2)\right] \label{del-rho}
\end{equation}
to the lowest order in $\Omega/D$. Equation~(\ref{del-rho}) can be solved by standard methods, which are straight forward but lengthy. We give the solution below.
\begin{eqnarray}
 \delta\rho(x)&=&-\Omega\frac{\Omega}{D}\left[\frac{k_1x^3}{3}+\frac{k_2x^2}{2}+k_3x\right]\nonumber \\&+&k^\prime\frac{\Omega}{D}\ln \left|x+\frac{2\rho_0-1}{\Omega}\right| + k_0.\label{delta-rho-basic}
\end{eqnarray}
Clearly, $\delta\rho(x)$ depends linearly on $\Omega/D$, and vanishes, as it should, when $\Omega/D$ vanishes. Here, $k_1,k_2,k_3,k^\prime$ are constants given by
\begin{eqnarray}
 &&k_1=-\frac{1}{12},\\
 &&k_2=-\frac{2\rho_0-1}{6\Omega},\\
 &&k_3=\frac{1}{\Omega} \left[\frac{\Omega}{12}+\frac{(2\rho_0-1)}{4}+\frac{(2\rho_0-1)^2}{6\Omega}\right],\\
 &&k^\prime=(2\rho_0-1)k_3.
\end{eqnarray}
Unsurprisingly, $\delta \rho(x)$ depends on $\rho_0$, which in turn is fixed by the boundary condition on $\rho_\text{LK}(x)$. We first focus on the LD phase. The constant of integration $k_0$ can be obtained by using the boundary conditions  $\delta\rho(x)=0$ at $x=0$. Now,
using the boundary condition at $x=0$, $\rho(0)=\alpha=\rho_\text{LK}(0)$ (which means $\delta\rho(x)=0$ at $x=0$, as we have used), we get $\rho_0=\alpha$. Then using (\ref{del-rho-def}) we obtain
\begin{eqnarray}
 &&\rho_\text{LD}(x)=\alpha +\frac{\Omega x}{2}-\Omega\frac{\Omega}{D}\left[\frac{k_1x^3}{3}+\frac{k_2x^2}{2}+k_3x \right]\nonumber \\&+&\frac{\Omega k^\prime}{D}\ln \left|x+\frac{2\alpha-1}{\Omega}\right|-\frac{\Omega k'}{D}\ln \left|\frac{2\alpha-1}{\Omega}\right|.\label{rho-ld-final}
 \end{eqnarray}
 As discussed above, corresponding to $\rho_\text{LD}(x)$ as given in (\ref{rho-ld-final}), the SEP density is given by $c_-(x)$, where
\begin{equation}
 c_-(x)= \frac{1}{2} + \frac{\Omega}{D}\left[\frac{\Omega}{12}(x-x^3)+\frac{1}{2}(\alpha-\frac{1}{2})(x-x^2)\right].\label{c-minus}
\end{equation}

Likewise, we can obtain $\rho_\text{HD}(x)$ by using the boundary condition at $x=1$, $\rho(1)=1-\beta=\rho_\text{LK}(1),\,\delta \rho(1)=0$. We get
 \begin{eqnarray}
 &&\rho_\text{HD}(x)=1-\beta+\frac{\Omega}{2}(x-1)\nonumber \\&-&\Omega\frac{\Omega}{D}\left[\frac{k_1}{3}(x^3-1)+\frac{k_2}{2}(x^2-1)+k_3(x-1)\right]\nonumber  \\
 &+&\frac{\Omega k^\prime}{D}\ln \left|x-1+\frac{1-2\beta}{\Omega} \right|-\frac{\Omega k'}{D}\ln \left|\frac{1-2\beta}{\Omega} \right|. \label{rho-hd-final}
\end{eqnarray}
Then corresponding to $\rho_\text{HD}(x)$ as given in (\ref{rho-hd-final}), the SEP density is $c_+(x)$ given by
\begin{equation}
 c_+(x)= \frac{1}{2} + \frac{\Omega}{D}\left[\frac{\Omega}{12}(x-x^3)+\frac{1}{2}\left(1-\beta-\frac{\Omega}{2}-\frac{1}{2}\right)(x-x^2)\right].\label{c-plus}
\end{equation}
Notice that in addition to the explicitly $x$-dependent solutions $\rho_\text{LD}(x)$ and $\rho_\text{HD}(x)$ above, the MFT equations (\ref{rho-mft}) and (\ref{c-mft}) also admit spatially uniform solutions $\rho=1/2,\,c=1/2$; $\rho=1/2$ obviously corresponds to the MC phase of the TASEP.
With these solutions for the steady state densities $\rho(x)$ and $c(x)$ phase diagrams for both the TASEP and SEP lanes can be constructed in the $\alpha-\beta$ plane. Since for large $D$, i.e., for small $\Omega/D$, we only expect small modifications of $\rho(x)$ from $\rho_\text{LK}(x)$ and of $c(x)$ from 1/2 in the bulk, we expect the TASEP phase diagram to be close to the one obtained in the LK-TASEP problem~\cite{frey-lktasep}, albeit with an equal attachment-detachment rate $\Omega'\equiv \Omega/2$.  



In our MCS studies with $D=1.0$ and $\Omega=0.3$ ($\Omega/D=0.3 < 1$), we find the so-called ``pure phases'' of TASEP (albeit generally space-dependent), {\em viz.}, LD, HD and MC phases and also detect the ``mixed phases'', e.g., LD-MC, HD-MC, LD-MC-HD and LD-HD phases. In these mixed phases, part of $\rho(x)$ is in one of the phases, and the remaining part is in another phase. The SEP density profiles may be characterised as follows. We define an average SEP density $\overline c$ via
\begin{equation}
 \overline c\equiv \int_0^1 c(x) dx.\label{mean-sep}
\end{equation}
Clearly, $\overline c>,<$ or $=1/2$ would imply excess, deficit and neutral phases. Furthermore, as we have discussed above, $c(x)$ tends to follow $\rho(x)$ in the bulk, although partially for non-zero $\Omega/D$. This implies, as our MCS studies on the SEP density profile reveal, $c(x)-1/2$ can cross zero in the bulk either once, or none at all, or remain zero over a finite extent of $x$.

We present an MFT phase diagram in Fig.~\ref{d10-phase-tasep} of the TASEP lane and an MFT phase diagram in Fig.~\ref{d10-phase-sep} of the SEP lane for $D=1.0$ and $\Omega=0.3$. We discuss how to obtain the phases and the corresponding phase boundaries between the phases in the MFT. Notice that both $\rho_\text{LD}(x)$ and $\rho_\text{HD}(x)$ are growing solutions of $x$. However, unlike in Ref.~\cite{frey-lktasep} they do not grow linearly with $x$; there are (small) nonlinear modifications to the linear growth profiles, which are due to the (large but) finite diffusivity in the SEP channel. Although there is no strict particle conservation in the bulk of the TASEP lane due to the attachment-detachment events, particle current is conserved {\em locally}, as locally the rate of particle nonconserving attachment-detachment events actually vanish in the thermodynamic limit~\cite{frey-lktasep}. As in the LK-TASEP model, steady state current here in the TASEP lane is space-dependent but continuous. Corresponding to the steady state densities $\rho_\text{LD}(x),\rho_\text{HD}(x)$ and 1/2, we define currents
\begin{eqnarray}
 j_\text{LD}(x)&=&\rho_\text{LD}(x) (1-\rho_\text{LD}(x)),\label{curr1}\\
 j_\text{HD}(x)&=&\rho_\text{HD}(x) (1-\rho_\text{HD}(x)),\label{curr2}\\
 j_\text{MC}(x)&=&\frac{1}{4} \label{curr3}.
\end{eqnarray}
Using the above expressions of the currents and their continuity across various phase boundaries~\cite{frey-lktasep}, we determine the location of the phase boundaries in the $\alpha-\beta$ plane. We set $j_\text{LD}(x_\alpha)=1/4$, equivalently $\rho_\text{LD}(x_\alpha)=1/2$, where $x_\alpha$ is the coordinate separating the LD phase from the MC phase, i.e., $\rho (x<x_\alpha)=\rho_\text{LD}(x<x_\alpha)<1/2$. Similarly, we set $j_\text{HD}(x_\beta)=1/4$, equivalently $\rho_\text{HD}(x_\beta)=1/2$, where $x_\beta$ is the coordinate separating the HD phase from the MC phase, i.e., $\rho (x>x_\beta)=\rho_\text{HD}(x>x_\beta)>1/2$. Depending upon the relative positions of $x_\alpha$ and $x_\beta$, various different density profiles emerge that we list briefly. (i) $x_\beta>x_\alpha\geq 1$ means the LD phase, (ii) $x_\beta>1$, $0<x_\alpha<1$ means the mixed LD-MC phase, with $\rho(x)<1/2$ for $0\leq x<x_\alpha$ and $\rho(x)=1/2$ for $x_\alpha<x<1$, (iii) $0<x_\alpha<x_\beta<1$ gives a three-phase coexistence with $\rho(x)<1/2$ for $0\leq x<x_\alpha$, $\rho(x)=1/2$ for $x_\alpha<x<x_\beta$ and $\rho(x)>1/2$ for $x_\beta<x\leq 1$. Further, $x_\alpha<0$, $0<x_\beta<1$ gives the HD-MC phase. It is also possible to have $x_\alpha>x_\beta$, whence one actually has the LD-HD phase with a domain wall at $x_w$.The position $x_w$ may be obtained from the condition $\rho_\text{LD}(x_w)+\rho_\text{HD}(x_w)=1$. Since this condition gives a unique solution for $x_w$, the domain wall in question is a {\em localised domain wall} or LDW located at $x_w$ with $0<x_w<1$.

The various phase boundaries in the $\alpha-\beta$ plane may be obtained in terms of the conditions on the densities as follows: setting (i) $\rho_\text{LD}(x_\alpha=1)=1/2$ gives the phase boundary between the LD and LD-MC phases, (ii) $\rho_\text{HD}(x_\beta=0)=1/2$, gives the phase boundary between the HD and HD-MC phases, (iii) $\rho_\text{LD}(x_\alpha=0)=1/2$ gives the phase boundary between the LD-MC and MC phases, (iv) $\rho_\text{HD}(x_\beta=1)=1/2$ gives the phase boundary between the HD-MC and MC phases, (v) $\rho_\text{LD}(x_w=1)+\rho_\text{HD}(x_w=1)=1$ gives the boundary between the LD and LD-HD phases (since this condition means that the domain wall is just at the right boundary $x=1$), (vi) $\rho_\text{LD}(x_w=0)+\rho_\text{HD}(x_w=0)=1$ gives the boundary between the LD-HD and HD phases (since this condition means that the domain wall is just at the left boundary $x=0$), (vii)  $\rho_\text{LD}(x_w=x_\alpha=x_\beta)+\rho_\text{HD}(x_w=x_\alpha=x_\beta)=1$ together with $\rho_\text{LD}(x_\alpha=x_\beta)=\rho_\text{HD}(x_\alpha=x_\beta)=1/2$ gives the boundary between the LD-HD and LD-HD-MC phases. These two conditions ensure that on the associated phase boundary, the size of the MC phase given by $x_\beta-x_\alpha$ just vanishes, indicating the threshold of the three-phase coexistence. The above-listed conditions are formally equivalent to solving for $x_\alpha,x_\beta,x_w$ and set conditions on them, as done in Ref.~\cite{frey-lktasep}. However, in Ref.~\cite{frey-lktasep} due to the linear dependence of the solutions of $\rho(x)$ on $x$, it was possible to explicitly solve for $x_\alpha,x_\beta,x_w$. The far more complex $x$-dependence of $\rho_\text{LD}(x)$ and $\rho_\text{HD}(x)$ rules out explicitly solving for $x_\alpha,x_\beta,x_w$. Instead, we obtain the phase boundaries by means of drawing contour plots in the  $\alpha$-$\beta$ plane for given values of $D$ and $\Omega$ corresponding to the conditions on the TASEP densities listed above; see the phase diagram in Fig.~\ref{d10-phase-tasep}. Notice that notwithstanding the far more complex $x$-dependence of $\rho_\text{LD}(x)$ and $\rho_\text{HD}(x)$, the phase diagram in Fig.~\ref{d10-phase-tasep} have straight lines parallel to either the $\alpha$- or $\beta$-axis as the phase boundaries between the LD and LD-MC phases, LD-MC and MC phases, HD and HD-MC phases, and MC and HD-MC phases. This is because $\rho_\text{LD}(x)$ and $\rho_\text{HD}(x)$ are independent, respectively, of $\beta$ and $\alpha$. This explains these phase boundaries. It is also clear that the phase diagram is invariant under the particle-hole symmetry. This may be seen by exchanging the $\alpha$ and $\beta$-axes and redrawing the phase boundaries. The resulting phase diagram is identical to that in Fig.~\ref{d10-phase-tasep}.

\begin{figure}[htb]
 \includegraphics[width=0.9\columnwidth]{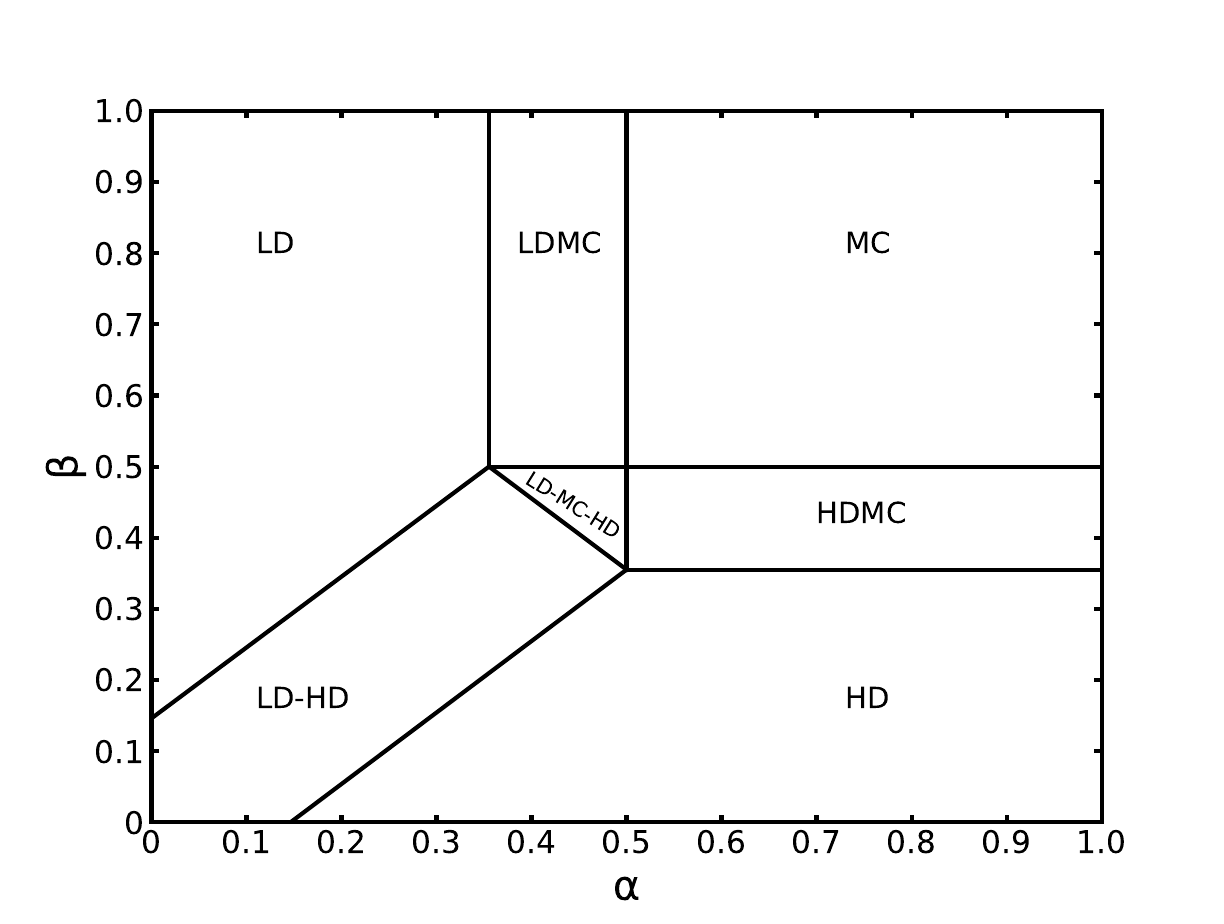}
 \caption{Mean-field TASEP phase diagram in the $\alpha-\beta$ plane with $D=1$, $\Omega=0.3$.}\label{d10-phase-tasep}
\end{figure}

\begin{figure}[htb]
 \includegraphics[width=0.9\columnwidth]{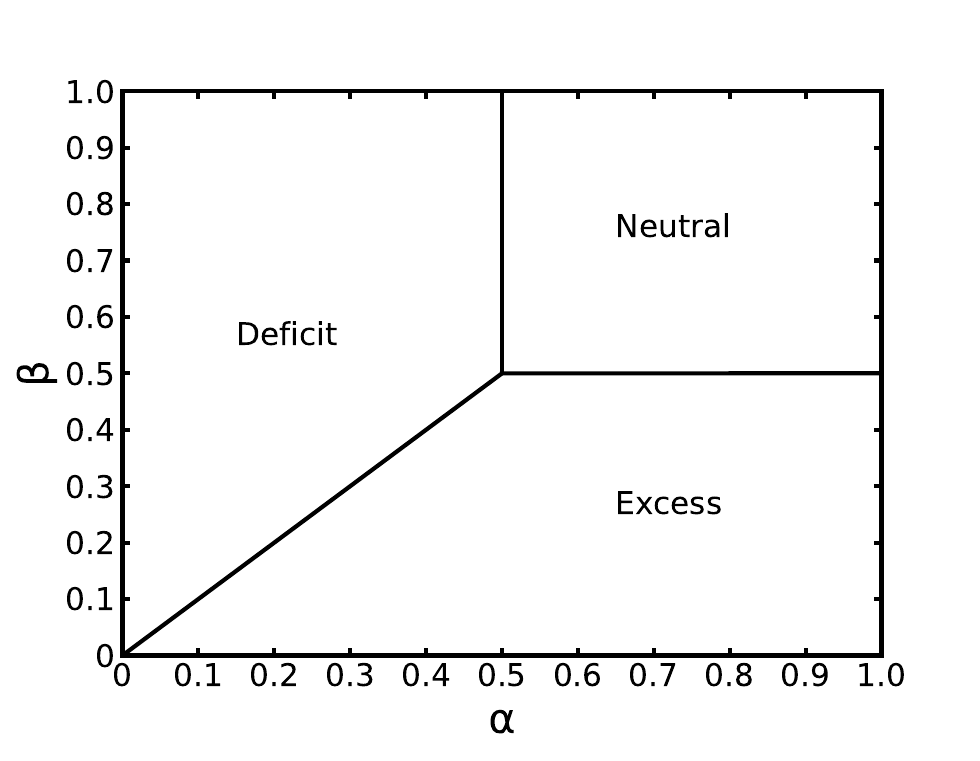}\\
 \caption{Mean-field SEP phase diagram in the $\alpha-\beta$ plane with $D=1$, $\Omega=0.3$.}\label{d10-phase-sep}
\end{figure}

An MFT phase diagram of the SEP lane for $\Omega=0.3, D=1$ is shown in Fig.~\ref{d10-phase-sep}. Phase space regions with excess, deficit and neutral particle numbers in SEP are shown, which are characterised by the mean SEP density $\overline c$ [see Eq.~(\ref{mean-sep}) above]. Due to the tendency of the SEP density $c(x)$ to follow the TASEP density $\rho(x)$ in the steady state, the pure LD, HD and MC phases in the TASEP lane correspond to deficit, excess and neutral particles. Furthermore, the quantity $c(x)-1/2$ is not zero in the bulk in some of the phase space regions of the SEP which correspond to the pure LD and HD phases  in the TASEP lane. There are however regions in the SEP phase diagram, corresponding to the LD-HD phases in the TASEP lane, where $c(x)-1/2$ crosses zero once in the bulk. In the remaining regions of the SEP phase diagram, $c(x)-1/2$ does not cross zero, but remains at zero in the whole or part of the bulk. These regions correspond to the pure MC phase, or mixed LD-MC or HD-MC or LD-MC-HD phases in the TASEP lane. The average steady state SEP density when the TASEP in its LD-MC (HD-MC) phase is less (more) than 1/2, implying that the SEP is in its deficit (excess) particle phase. When the TASEP is in its LD-HD phase with an LDW at $x=x_w$, the quantity $c(x)-1/2<0$ for $0\leq x< x_w$ and $c(x)-1/2>0$ for $x_w< x\leq 1$. When $x_w=1/2$, the LDW is located at the mid-point of the TASEP channel, when happens on the line $\alpha =\beta$. Specifically on this line,  $c(x)-1/2<0$ for $0\leq x< 1/2$ and $c(x)-1/2>0$ for $1/2< x\leq 1$, giving $\alpha=\beta$ as the phase boundary between the deficit and excess particle phases of the SEP, when the TASEP is in its LD-HD phase. When the TASEP is in its LD-MC-HD phase, $c(x)-1/2<0$ for $0\leq x<x_\alpha$, $c(x)-1/2=0$ for $x_\alpha<x<x_\beta$ and  $c(x)-1/2>0$ for $x_\beta<x\leq 1$. Furthermore, symmetry of the model about the line $\alpha=\beta$ (which has its origin in the particle-hole symmetry of the model; see discussions above) ensures that  $\alpha=\beta$ continues to be the boundary between the deficit and excess particle phases of TASEP. This line terminates at the multicritical point $\alpha=1/2=\beta$, whence it meets the neutral phase. These discussions suggest that ${\cal O}_c\equiv \overline c-1/2$ may be used as an order parameter to delineate and distinguish the different phases in the SEP. Indeed, in the neutral phase ${\cal O}_c=0$, in the deficit phase ${\cal O}_c<0$ and in the excess phase ${\cal O}_c>0$.  All the three SEP phase boundaries are second order in nature, which meet at the multicritical point (1/2,1/2).

We verify the accuracy of the above MFT by comparing with the numerical results on the steady state density profiles obtained from our extensive MCS studies with $D=1,\,\Omega=0.3$. See Fig.~\ref{d-10-tasep} for representative plots of $\rho(x)$ as a function of $x$ in different phases of the TASEP with $D=1,\,\Omega=0.3$. Analytical MFT and numerical MCS results are superposed. For MFT, we have used $\rho(x)=\rho_\text{LD}(x)$ [Eq.~(\ref{rho-ld-final})] in the LD phase of the TASEP,  $\rho(x)=\rho_\text{HD}(x)$ [Eq.~(\ref{rho-hd-final})] in the HD phase, and $\rho(x)=1/2$ in the MC phase. We have presented our results on the corresponding SEP density profile $c(x)$ in  Fig.~\ref{d-10-sep}. Both analytical MFT and numerical MCS results are shown. Reasonably good agreements are found between the MFT and MCS results. For MFT solution of the SEP density, we have used $c(x)=c_-(x)$ [Eq.~(\ref{c-minus})] for $c(x)<1/2$, corresponding to the TASEP in its LD phase, and $c(x)=c_+(x)$ [Eq.~(\ref{c-plus})] for $c(x)>1/2$, corresponding to the TASEP in its HD phase. 
{Notice that the quantitative agreement of the MFT solutions for the SEP density $c(x)$ with the corresponding MCS results 
is good when the TASEP is in its LD or HD phases, but less so when the TASEP in its LD-HD phase; see Fig.~\ref{d-10-sep}. We believe this is due to the fact that near the location of an LDW in TASEP, $c(x)$ has a strong space dependence, suggesting importance of retaining the higher order corrections in the MFT solutions of $c(x)$. Nevertheless, qualitative agreement between the MFT and MCS solutions for $c(x)$ can be seen even in this case.} { In addition to the steady state densities $\rho(x)$ and $c(x)$, we have also plotted the corresponding steady currents: bulk TASEP current $j_\text{MCS}(x)=\rho(x)(1-\rho (x))$ and SEP current $ j_s(x)=-D\partial_x c(x)$, where we have used the respective MFT definitions of the currents, and have evaluated them from the corresponding MCS results for $\rho(x)$ and $c(x)$; see Fig.~\ref{d-10-tasep} and Fig.~\ref{d-10-sep}. Due to the very weak $x$-dependence, the SEP current $j_s(x)$ having small values is shown in the insets of Fig.~\ref{d-10-sep} for better resolution. The numerical smallness of $j_s(x)$ can be understood as follow. We note that the SEP current $j_s(x)$ in an isolated unbiased SEP vanishes. In the present problem, it is non-zero but small for large $D$, since in that limit, the steady state SEP density $c(x)$ should be close to that of an isolated unbiased SEP, which is 1/2. This gives $j_s(x)=-D\partial_x c(x)$ to be very small and nearly flat. }

\begin{figure}[htb]
 \includegraphics[width=0.85\columnwidth]{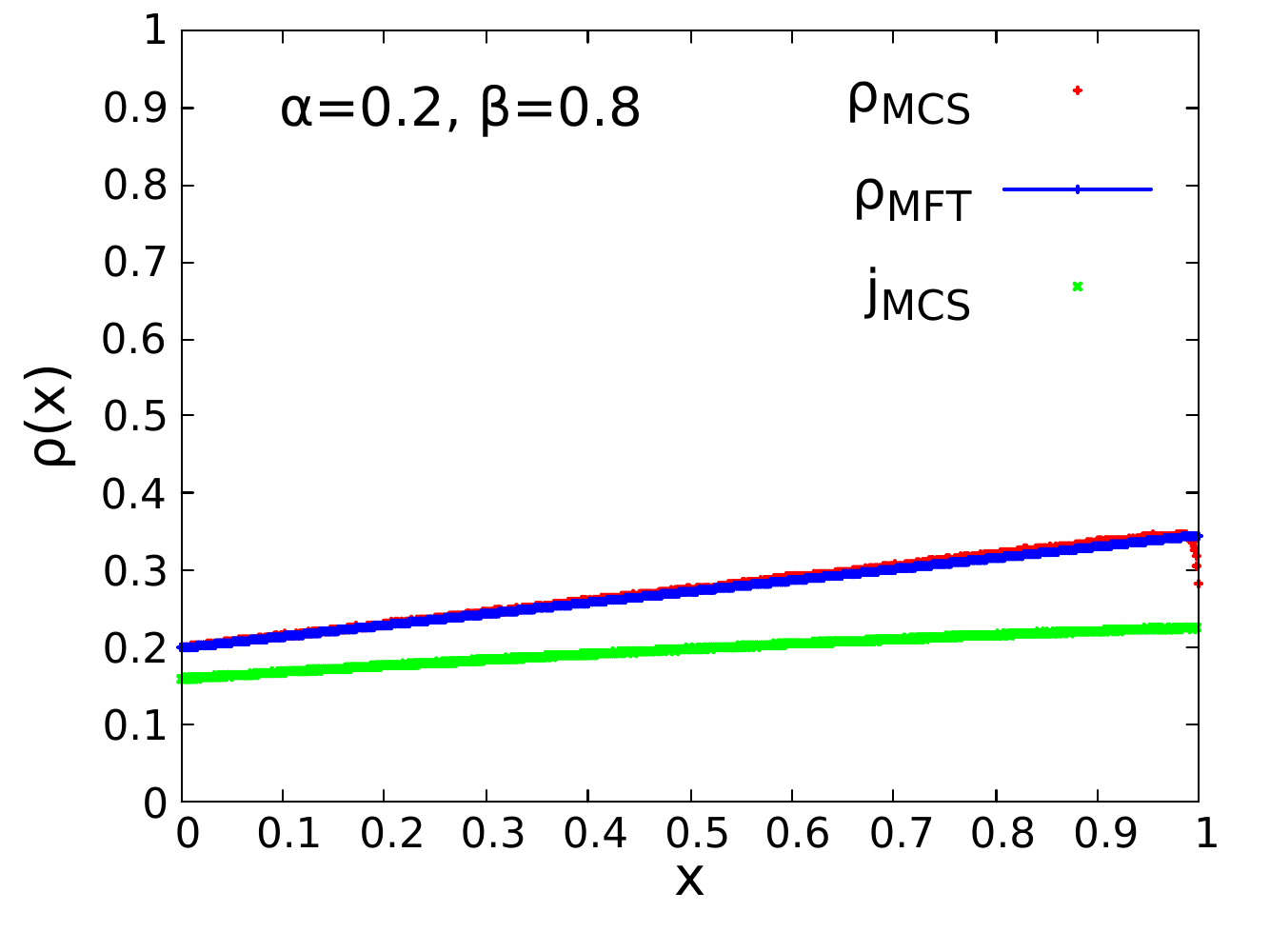}\hfill
 \includegraphics[width=0.85\columnwidth]{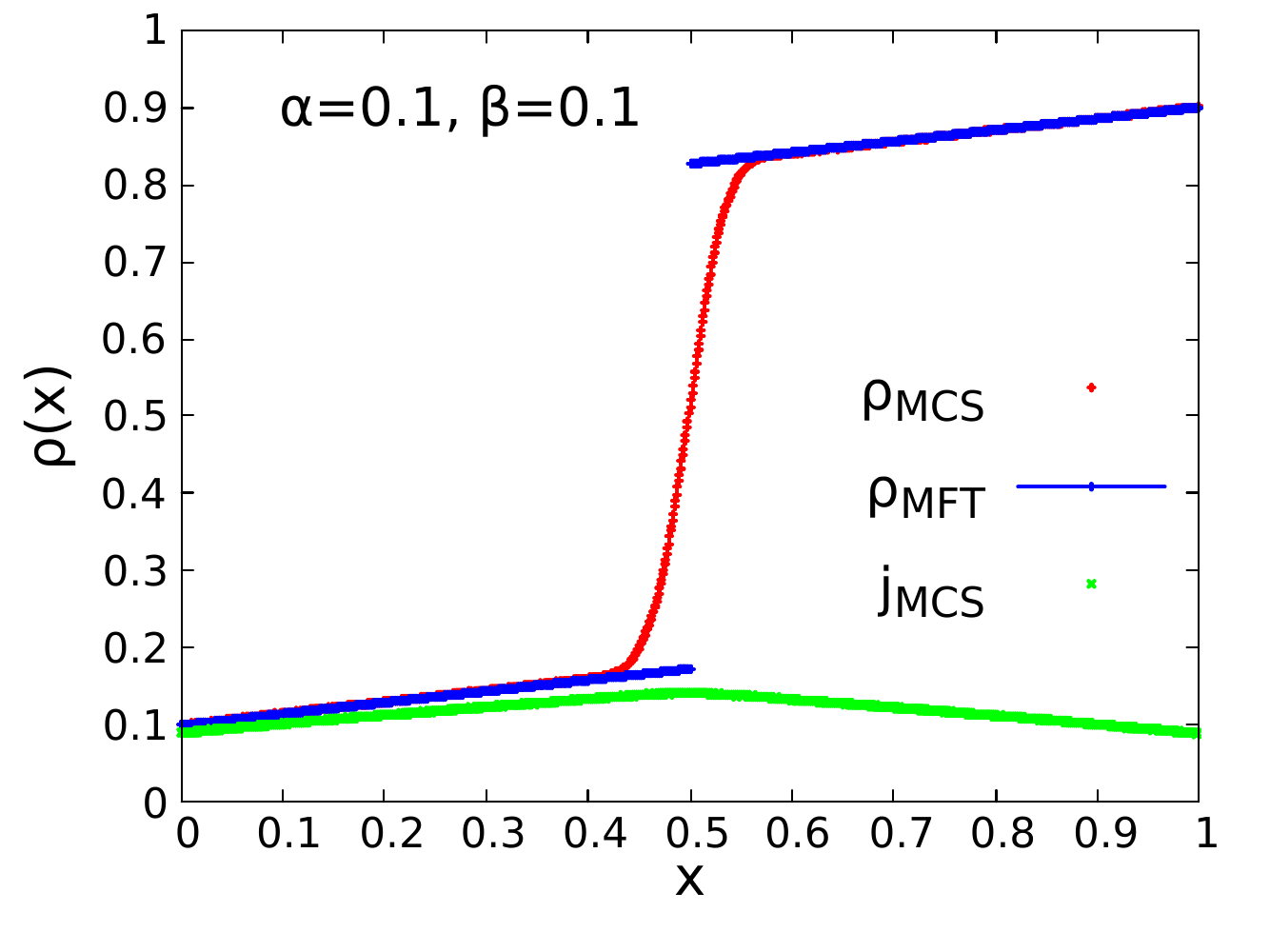}\\
 \includegraphics[width=0.85\columnwidth]{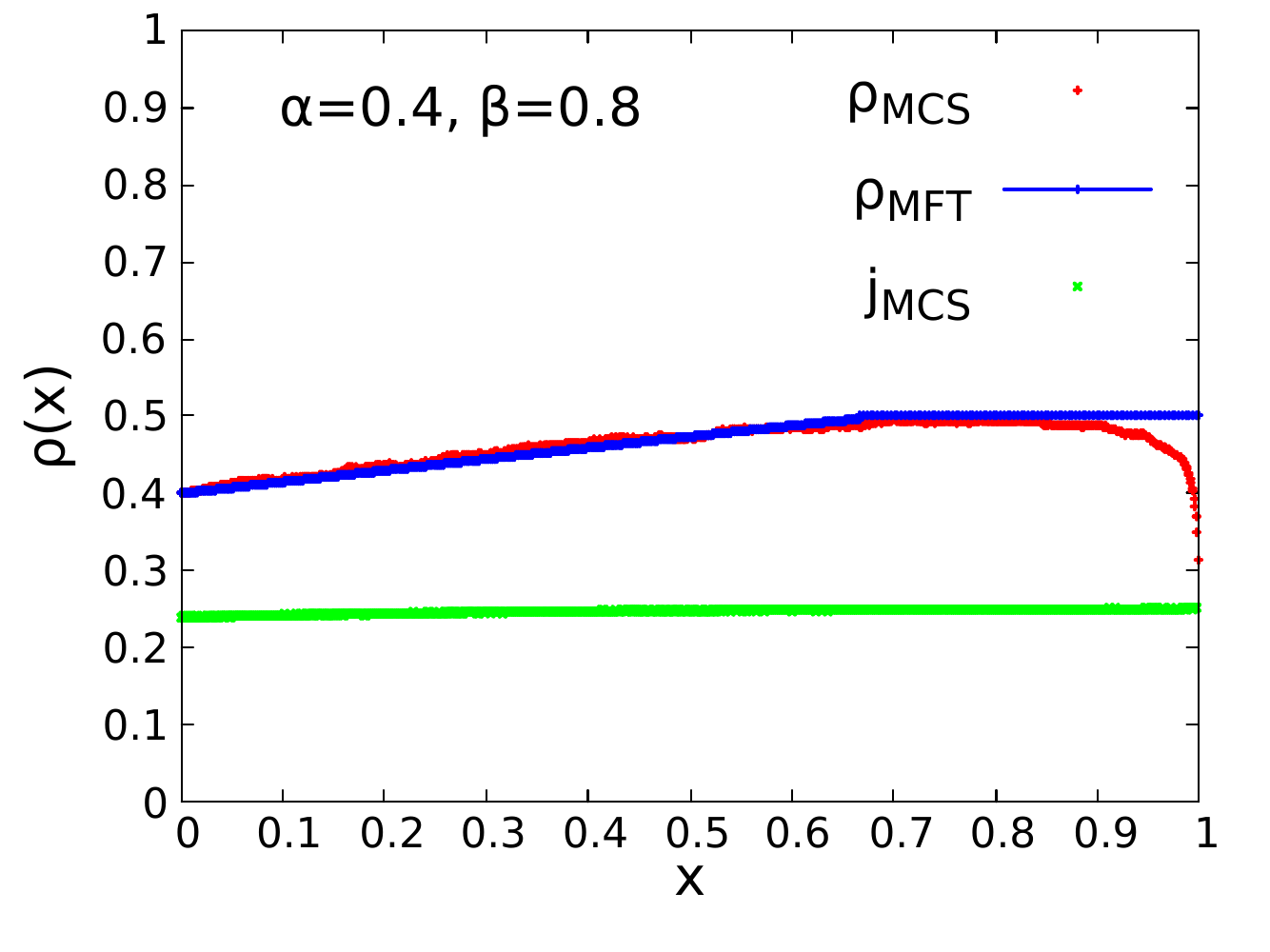}
 \caption{Steady state density $\rho(x)$ in the TASEP lane in the LD (top), LD-HD (middle) and LD-MC (bottom) phases with $D=1,\,\Omega=0.3$. MFT (blue line) and MCS (red points) results are shown. For MFT, we have used $\rho(x)=\rho_\text{LD}(x)$ [Eq.~(\ref{rho-ld-final})] in the LD phase of the TASEP, $\rho(x)=\rho_\text{HD}(x)$ [Eq.~(\ref{rho-hd-final})] in the HD phase and $\rho(x)=1/2$ in the MC phase (see text). { The corresponding steady state TASEP current $j_\text{MCS}(x)$, as obtained from the MCS solutions, are shown in green.} }\label{d-10-tasep}
\end{figure}

\begin{figure}[htb]
 \includegraphics[width=0.85\columnwidth]{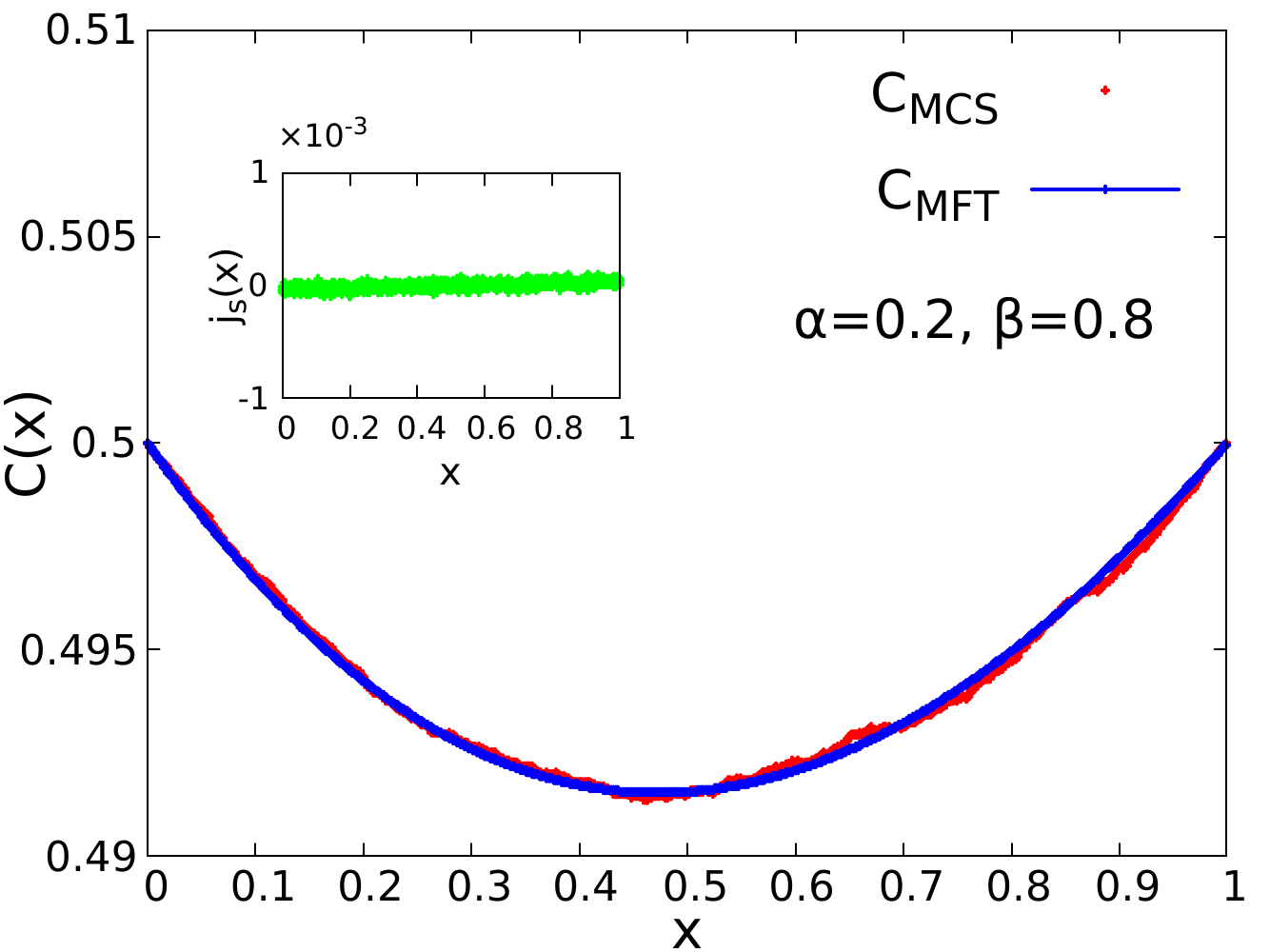}\hfill
 \includegraphics[width=0.85\columnwidth]{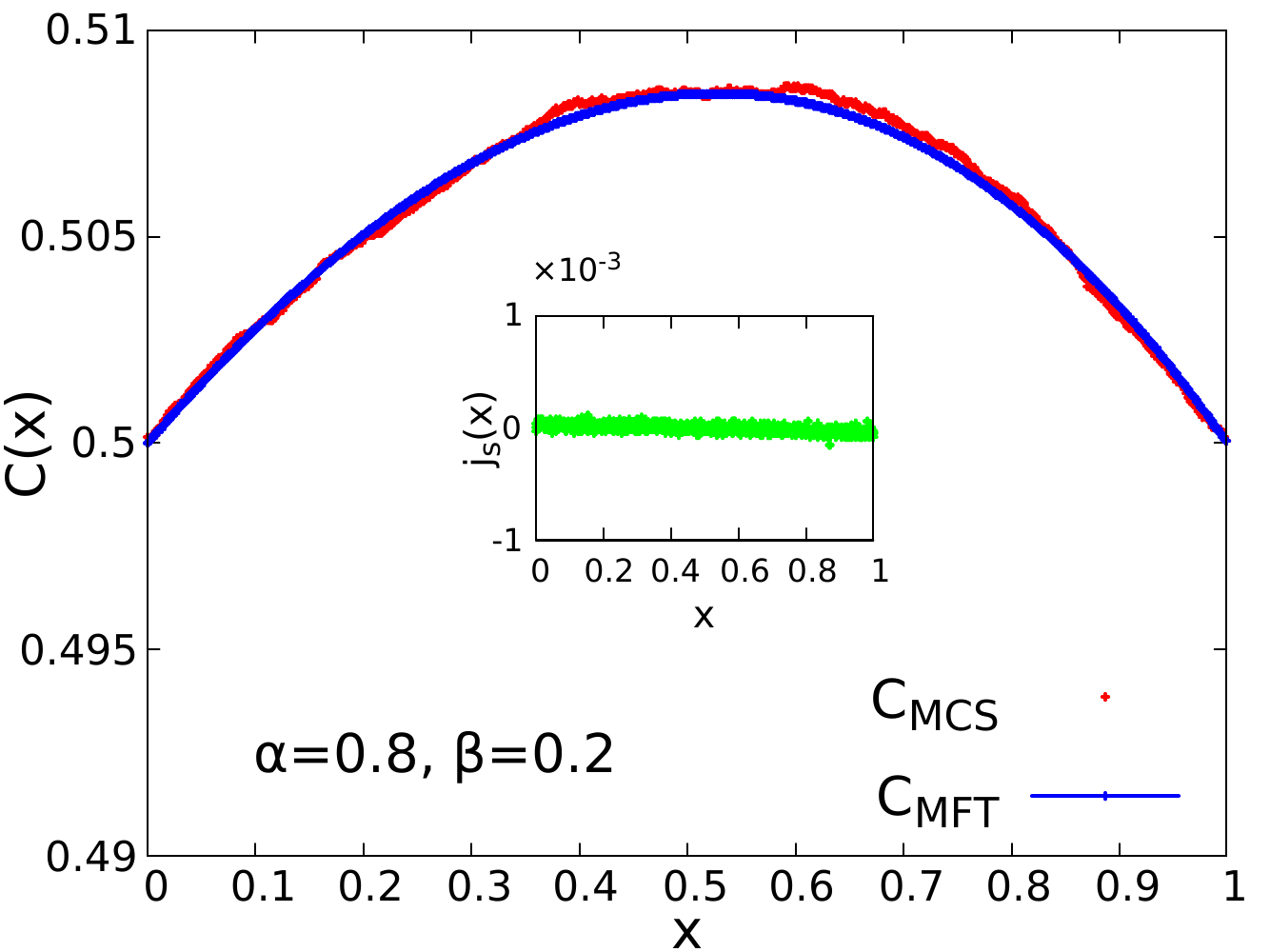}\\
 \includegraphics[width=0.85\columnwidth]{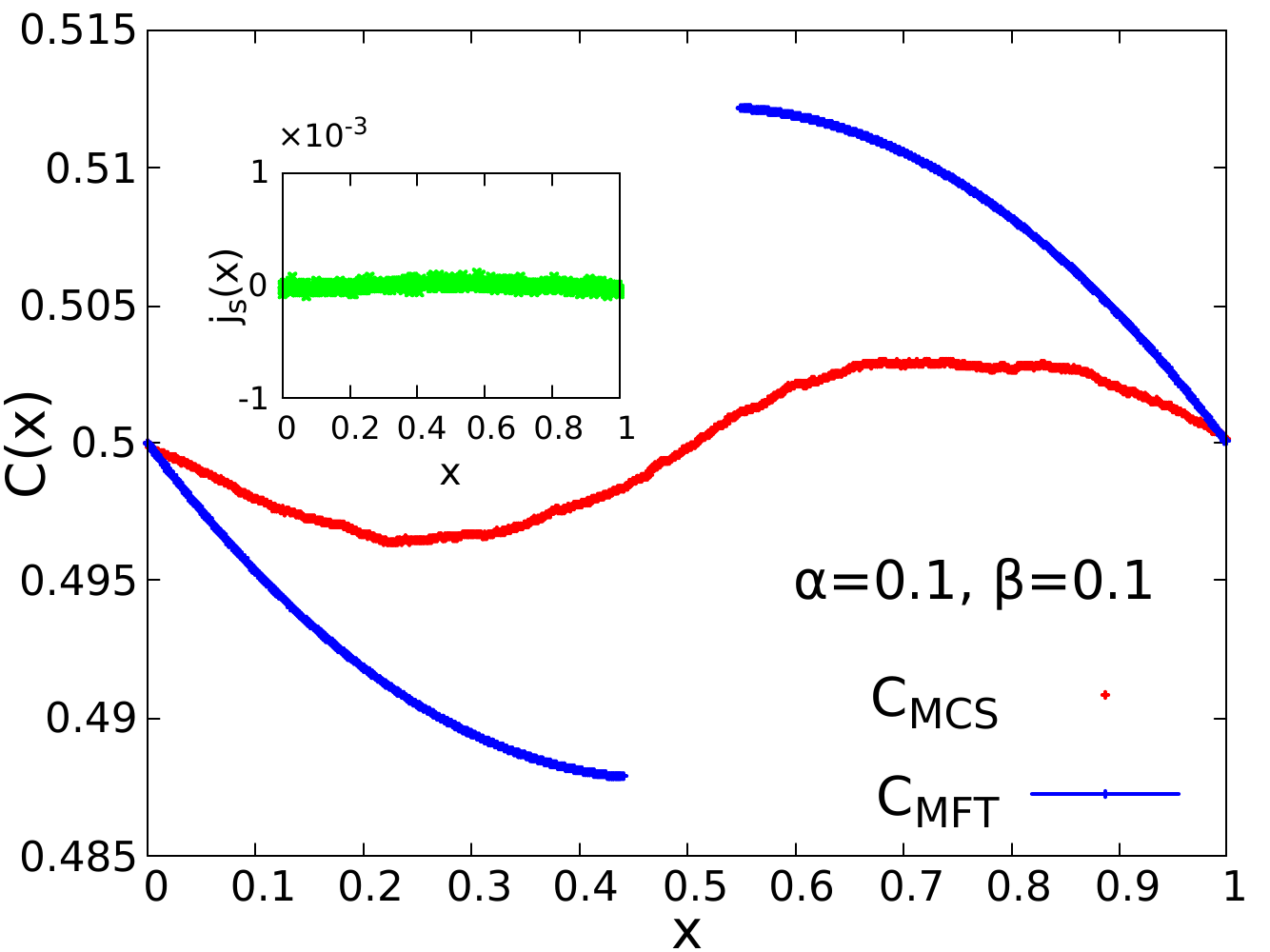}  
 \caption{Steady state density $c(x)$ in the SEP lane, when the TASEP lane is in its LD (top), HD (middle) and LD-HD (bottom) phases with $D=1,\,\Omega=0.3$. MFT (blue line) and MCS (red points) results are shown. For MFT, we have used $c(x)=c_-(x)$ [Eq.~(\ref{c-minus})] for $c(x)<1/2$, corresponding to the TASEP in its LD phase, and $c(x)=c_+(x)$ [Eq.~(\ref{c-plus})] for $c(x)>1/2$, corresponding to the TASEP in its HD phase (see text). { In the inset of each figure, the corresponding SEP current $j_s(x)$ is plotted as a function of $x$.}}\label{d-10-sep}
\end{figure}



\subsection{MFT for small $D$}

We now consider the solutions of the MFT equations (\ref{rho-mft}) and (\ref{c-mft}) for small values of $D$ with a fixed $\Omega$, i.e., $\Omega/D\gg 1$. As discussed above, for $\Omega/D\rightarrow \infty$, $\rho(x)$ reduces to $\rho_{T}(x)$ and $c(x)=\rho(x)$ in the bulk, where $\rho_T$ is the bulk steady state density in an open isolated TASEP. This solution for $c$ however does not match with the boundary condition except when $\rho=1/2=\rho_\text{MC}$. To impose the boundary conditions, for all other steady state solutions for $\rho(x)$, there are {\em two} boundary layers close to the two boundaries at $x=0,1$, ensuring $c(0)=1/2=c(1)$. These boundary layers are analogous to the boundary layers observed in the MCS studies on the steady state density profiles in an open TASEP.  For $\Omega/D$ large but finite, we expect small modifications   to this picture. 

To find the steady state densities in both the SEP and TASEP channels, we proceed as follows. We have already noted that the exchange of particles (subject to exclusion) between the TASEP and SEP channels maintains the overall particle number conservation in the combined bulk of the two lanes. This gives rise to the conservation law (\ref{mft-con}) above, which is a quadratic equation in $\rho(x)$ in terms of $J$ and other quantities. 
Now by using (\ref{mft-con}) and assuming small $D$, we write the two solutions of $\rho(x)$ in terms of $J$ and $c(x)$ as
\begin{eqnarray}
 \rho_\pm(x) &=& \frac{1}{2}\left[1\pm\sqrt{J-4D\partial_x c(x)}\right]\nonumber \\
 &\approx& \frac{1}{2}(1\pm\sqrt J)\mp \frac{D}{\sqrt J}\partial_x c(x).\label{rho-large-D}
\end{eqnarray}
Clearly, $\rho_+(x)>1/2$ and $\rho_-(x)<1/2$ in the bulk of the TASEP.  
We now use (\ref{rho-large-D}) to eliminate $\rho$ in (\ref{c-mft}) to obtain a single closed equation for $c(x)$:
\begin{equation}
 D\partial_x^2 c - A_\pm \partial_x c - \Omega [c(x) - B_\pm]=0,\label{c-eff-eq}
\end{equation}
where using $\rho_+ (\rho_-)$ for $\rho$ gives (\ref{c-eff-eq}) with $A_+,\,B_+\,(A_-,\,B_-)$, which depend on $J$. Here,
\begin{equation}
 A_\pm = \pm \frac{\Omega D}{\sqrt J_\pm},\, B_\pm=\frac{1}{2}(1 \pm \sqrt J_\pm).
\end{equation}
What is $J$ here? We note that in the limit of $D\rightarrow 0$, $\rho(x)\rightarrow \rho_T$, where $\rho_T$ is the MFT solution of the steady state density in an open TASEP. Thus in that limit, $J=J_-=(2\alpha-1)^2$, if $\rho_T=\rho_\text{LD}$, whereas $J=J_+=(2\beta-1)^2$ if $\rho_T=\rho_\text{HD}=1-\beta$. These considerations give, as expected,
\begin{eqnarray}
 \rho_-(x)=\alpha=\rho_\text{LD},\\
 \rho_+(x)=1-\beta=\rho_\text{HD},
\end{eqnarray}
when $D\rightarrow 0$, coinciding with the MFT solutions for an open TASEP.  

Solving (\ref{c-eff-eq}) we get two solutions for $c(x)$:
\begin{equation}
 c_\pm(x)=B_\pm + U^\pm_1\exp(\lambda^\pm_1 x)+U^\pm_2\exp(\lambda^\pm_2 x),\label{ceq-soln}
\end{equation}
where
\begin{eqnarray}
 \lambda_1^+&=&\frac{1}{2D}\bigg[A_++\sqrt{A_+^2+4D\Omega}\bigg],\\
 \lambda_2^+&=&\frac{1}{2D}\bigg[A_+-\sqrt{A_+^2+4D\Omega}\bigg]
\end{eqnarray}
corresponding to $\rho=\rho_+$, and
\begin{eqnarray}
 \lambda_1^-&=&\frac{1}{2D}\bigg[A_-+\sqrt{A_-^2+4D\Omega}\bigg],\\
 \lambda_2^-&=&\frac{1}{2D}\bigg[A_--\sqrt{A_-^2+4D\Omega}\bigg]
\end{eqnarray}
for $\rho=\rho_-$.
Here, $U^\pm_1,U^\pm_2$ are two sets of constants of integration, to be evaluated by using the two boundary conditions on $c$, {\em viz.}, $c(0)=1/2=c(1)$. As for $A_\pm,\,B_\pm$, use of $\rho_+(x)\; (\rho_-(x))$ as the solution for $\rho(x)$ correspond to the set $U^+_1,\,U^+_2\,(U^-_1,\,U^-_2)$. We find
\begin{eqnarray}
 U_1^\pm&=& \frac{1-\exp(\lambda_2^\pm)}{\exp(\lambda_1^\pm)-\exp(\lambda_2^\pm)}\bigg(\frac{1}{2}-B_\pm\bigg),\\
 U_2^\pm&=& \frac{\exp(\lambda_1^\pm)-1}{\exp(\lambda_1^\pm)-\exp(\lambda_2^\pm)}\bigg(\frac{1}{2}-B_\pm\bigg).
\end{eqnarray}
Evaluation of the constants allow us to find $c_-(x)$ and $c_+(x)$, which in turn yield $\rho_\text{LD}(x)$ and $\rho_\text{HD}(x)$ respectively.

For finite but small $D$, we expect weak space dependence of $\rho_-(x)$ and $\rho_+(x)$, i.e., weakly deviating from the constant solutions of $\alpha$ and $1-\beta$ respectively in the bulk. For a finite but small $D$, identifying $\rho_-(x)<1/2$ with $\rho_\text{LD}(x)$ and $\rho_+(x)>1/2$ with $\rho_\text{HD}(x)$, we find 
\begin{eqnarray}
 \rho_\text{LD}(x)&=&\frac{1}{2}\bigg[1-\sqrt{J_\text{LD}-4D\partial_x c_-(x)}\bigg]<\frac{1}{2},\label{rho-ld-smallD}\\
 \rho_\text{HD}(x)&=&\frac{1}{2}\bigg[1+\sqrt{J_\text{HD}-4D\partial_x c_+(x)}\bigg]>\frac{1}{2} \label{rho-hd-smallD}
\end{eqnarray}
for small $D$.
In general, $J_\text{LD}$ and $J_\text{HD}$ should now include current contributions from the SEP channel; see Eq.~(\ref{mft-con}) above. When the TASEP lane is in its LD (HD) phase, its bulk solution and the associated current is controlled by the left (right) boundary condition. We then identify 
\begin{eqnarray}
 J_\text{LD}&=& (2\alpha-1)^2+ 4D\partial_x c_-(x)|_{x=0},\\
 J_\text{HD}&=&(2\beta-1)^2 + 4D\partial_x c_+(x)|_{x=1}.
\end{eqnarray}
Here, $c_-(x)$ and $c_+(x)$ are the two solutions [see the Eq.~(\ref{ceq-soln})] of  (\ref{c-eff-eq}), obtained by using $\rho(x)=\rho_-(x)$ and $\rho(x)=\rho_+(x)$ respectively.

Equations~(\ref{ceq-soln}), (\ref{rho-ld-smallD}) and (\ref{rho-hd-smallD}) provide the MFT solutions for $c(x)$ and $\rho(x)$ valid in the limit of small $D$. Notice that $J_\text{LD/HD}$ appears in $\rho_\text{LD/HD}(x)$. Thus knowledge of $J_\text{LD/HD}$ is necessary to evaluate $\rho_\text{LD/HD}(x)$. Now, $J_\text{LD/HD}$ depends upon $c(x)$ through its spatial derivative $\partial_x c(x)$ obtained at $x=0,1$. On the other hand, enumeration of  $c(x)$ requires $J_{\pm}$, because of the dependence of the constants $A_\pm, B_\pm$ etc on it. 
For simplicity while evaluating the currents, we approximate $J_{\pm}$ by dropping the contributions from the SEP current to it, rendering it calculable from the TASEP boundary conditions only. Thus, in this approximation $J_-=(2\alpha-1)^2,\, J_+=(2\beta-1)^2$. This is a reasonable approximation, since for small $D$, for which the current analysis holds, $c(x)$ is largely slaved to $\rho(x)$ in the bulk, making it rather weakly space-dependent, which in turn means the SEP current should be much smaller than the TASEP current. Lastly, notice that $\rho(x)=1/2=c(x)$ continue to be steady state solutions for both $c(x)$ and $\rho(x)$ in the bulk, even when $D$ does not vanish. Now $\rho(x)=1/2$ implies MC phase in TASEP, which means when the TASEP is in its MC phase, the SEP and TASEP densities should overlap in the bulk for any finite $D$.

Having found the steady states solutions of $\rho(x)$ and $c(x)$ for small $D$, we now obtain the phase diagrams for both the TASEP and SEP lanes. Since for small $D$, $\rho(x)$ varies weakly with $x$, we expect a TASEP phase diagram similar to that in an open TASEP, with most of the phase diagram being covered by regions with pure LD, HD or MC phases. However, due to the expected weak space dependence of $\rho(x)$ (as opposed to constant $\rho$ in pure open TASEPs), mixed phases other than the pure phases should also appear albeit over smaller ranges of $\alpha,\beta$, which should go to zero as $D\rightarrow 0$. We present an MFT phase diagram in Fig.~\ref{d01-phase-tasep} of the TASEP lane for $D=0.01$ and $\Omega=0.3$ (thus $\Omega/D=30\gg 1$). The principles to obtain the TASEP phase diagram are same as those discussed in the large $D$ case above. We use the continuity of the currents (\ref{curr1})-(\ref{curr3}) across the phase boundaries. As with $D>>1$, the rather complex $x$-dependence of $\rho(x)$ precludes explicit enumeration of $x_\alpha,x_\beta,x_w$. Instead, we use the conditions on the densities as listed in the previous Section, and then use contour plots in the $\alpha-\beta$ plane for fixed $D$ and $\Omega$ to obtain the phase boundaries.
\begin{figure}[htb]
 \includegraphics[width=0.9\columnwidth]{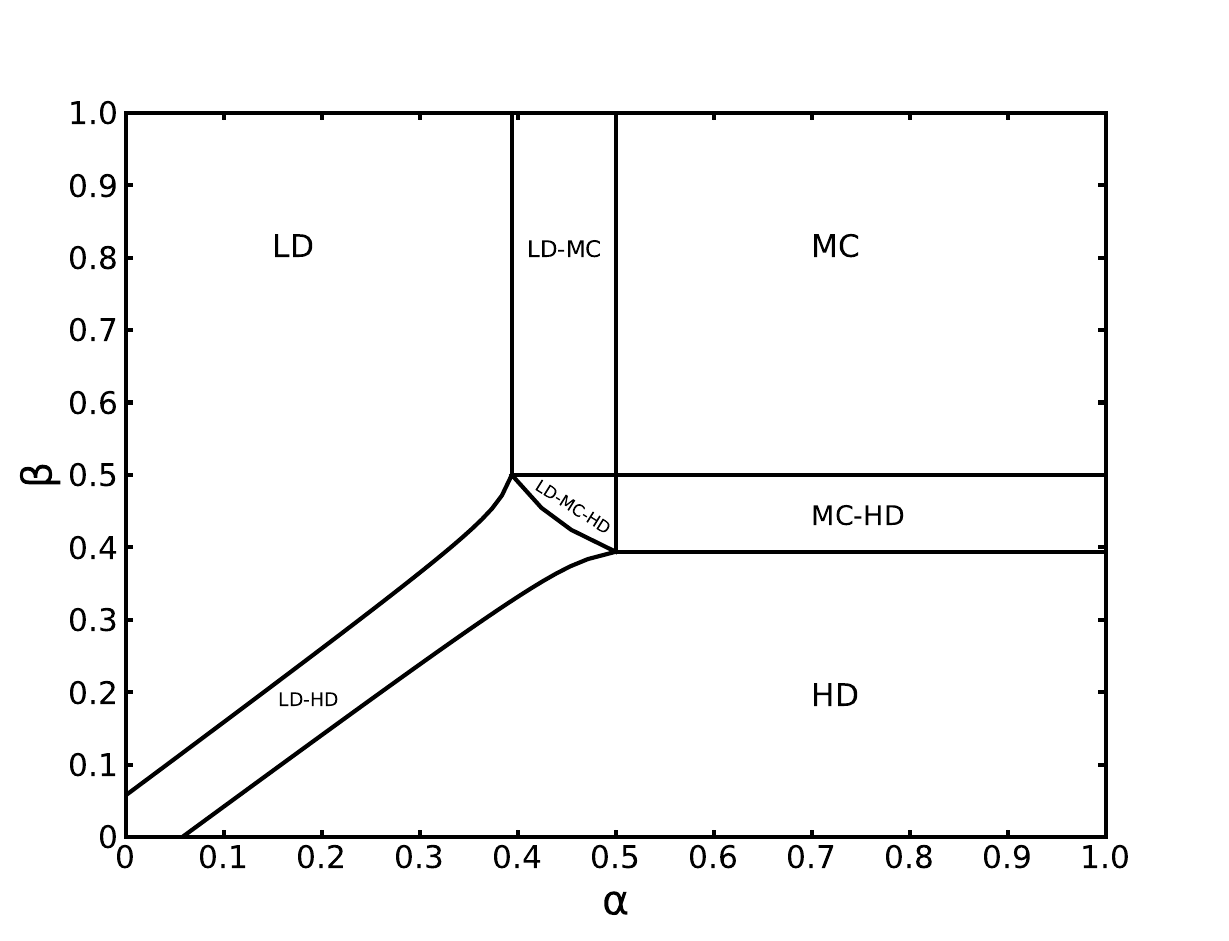}\\
 \caption{Mean-field TASEP phase diagrams with $D=0.01$, $\Omega=0.3$.}\label{d01-phase-tasep}
\end{figure}

The corresponding phase diagram of the SEP channel with $D=0.01$ and $\Omega=0.3$ is same as that given in Fig.~\ref{d10-phase-sep} with three second order phase boundaries meeting at a multicritical point located at $\alpha=1/2=\beta$. In fact, the SEP phase diagram is same for any finite value of $D$. This may be understood as follows. From the particle-hole symmetry of the model, the LD phase regions (covering both the pure LD phase and the LD parts of the mixed phases) must have the same area as that of the HD phase regions in the $\alpha-\beta$ plane of a TASEP phase diagram. Furthermore, these two regions are also {\em symmetrically} located on the two sides of the line $\alpha=\beta$. According to logic outlined above, these two regions correspond to the deficit and excess particle regions in a SEP phase diagram, which are also symmetrically located on the two sides of the line $\alpha=\beta$. The remaining region in a SEP phase diagram is the neutral region. Since these arguments hold for any finite $D$, the SEP phase diagram remains unchanged when $D$ varies. There is however one thing that changes, {\em viz.}, the amount of ``excess'' or ``deficit'' particles in the corresponding phase regions of TASEP. As the SEP gets increasingly slaved to the TASEP when $D$ is progressively reduced, for the same $\alpha,\beta$ the degree of ``excess'' or ``deficit'' rises   (assuming the TASEP is not in its MC phase), reaching the maximum for $D\rightarrow 0$. In the opposite extreme limit when $D\rightarrow \infty$, $c(x)\rightarrow 1/2$ for all $\alpha,\beta$, meaning the whole SEP phase diagram now has only the neutral phase. 

Plots of the steady state TASEP density $\rho(x)$ and SEP density $c(x)$ versus $x$ for $D=0.01$ and $\Omega=0.3$ are shown respectively in Fig.~\ref{d-01-tasep} and Fig.~\ref{d-01-sep}. Both analytical MFT and numerical MCS results are shown. For MFT solutions of TASEP, we have used $\rho(x)=\rho_\text{LD}(x)$ [Eq.~(\ref{rho-ld-smallD})] in the LD phase and $\rho(x)=\rho_\text{HD}(x)$ [Eq.~(\ref{rho-hd-smallD})] in the HD phase of the TASEP. In addition, the solutions $\rho(x)=1/2$ corresponds to the MC phase of the TASEP. For MFT solutions of SEP, we have used $c(x)=c_-(x)<1/2$ and $c(x)=c_+(x)>1/2$ as defined in Eq.~(\ref{ceq-soln})] above.  { As for the large $D$ case, we have again plotted $j_\text{MCS}(x)$ and $j_s(x)$ along side the corresponding density profiles;  see Fig.~\ref{d-01-tasep} and Fig.~\ref{d-01-sep}, with $j_s(x)$ being shown in the insets of Fig.~\ref{d-01-sep} due to its numerical smallness. Indeed in this case, $j_s(x)$ continues to be small, although for a reason different from the large $D$ case: While in this case, $c(x)$ has a substantial spatial variation, for a small $D$, $j_s(x)=-D\partial_x c$ continues to be small.}

\begin{figure}[htb]
 \includegraphics[width=0.85\columnwidth]{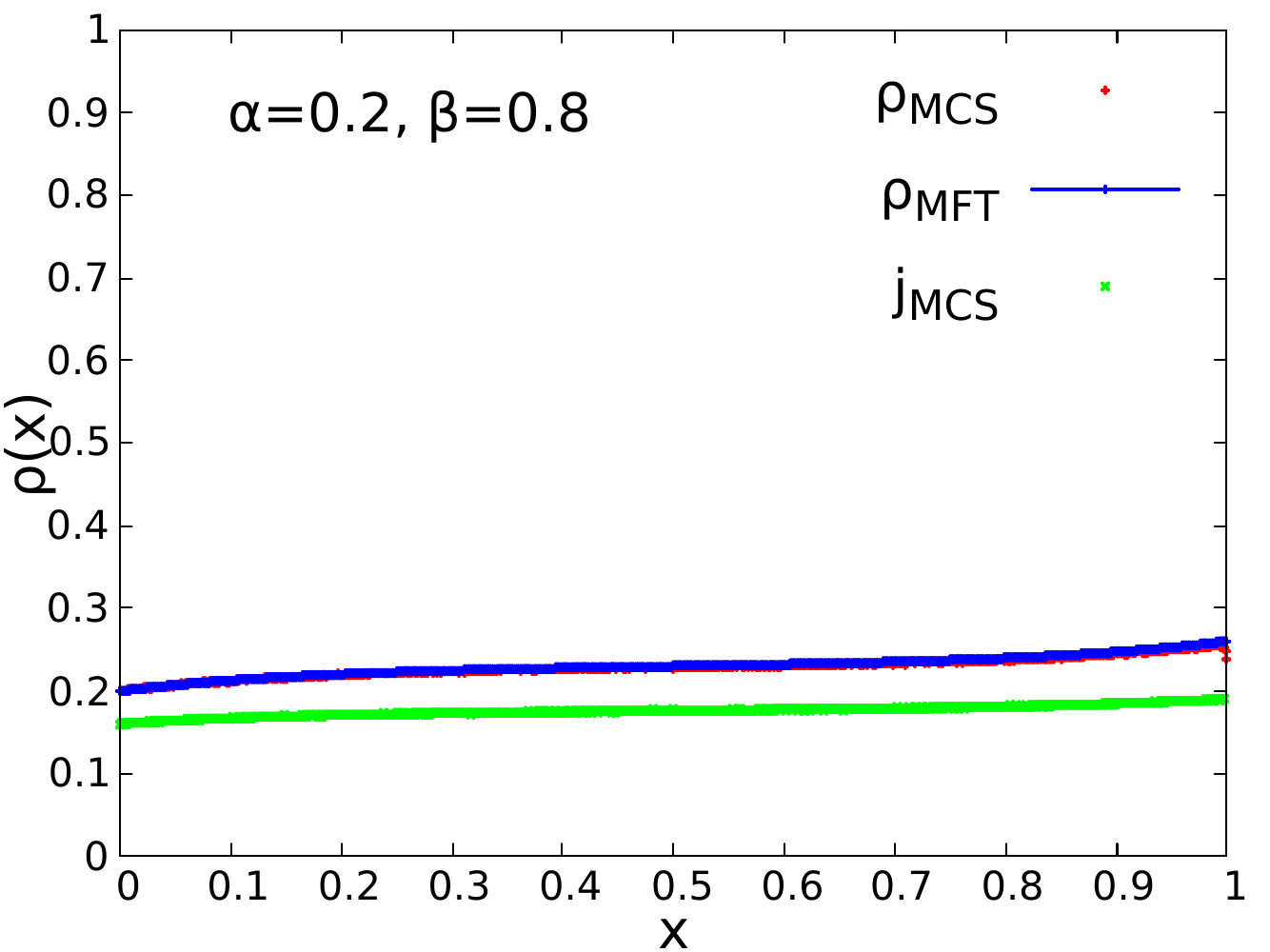}\\
 \includegraphics[width=0.85\columnwidth]{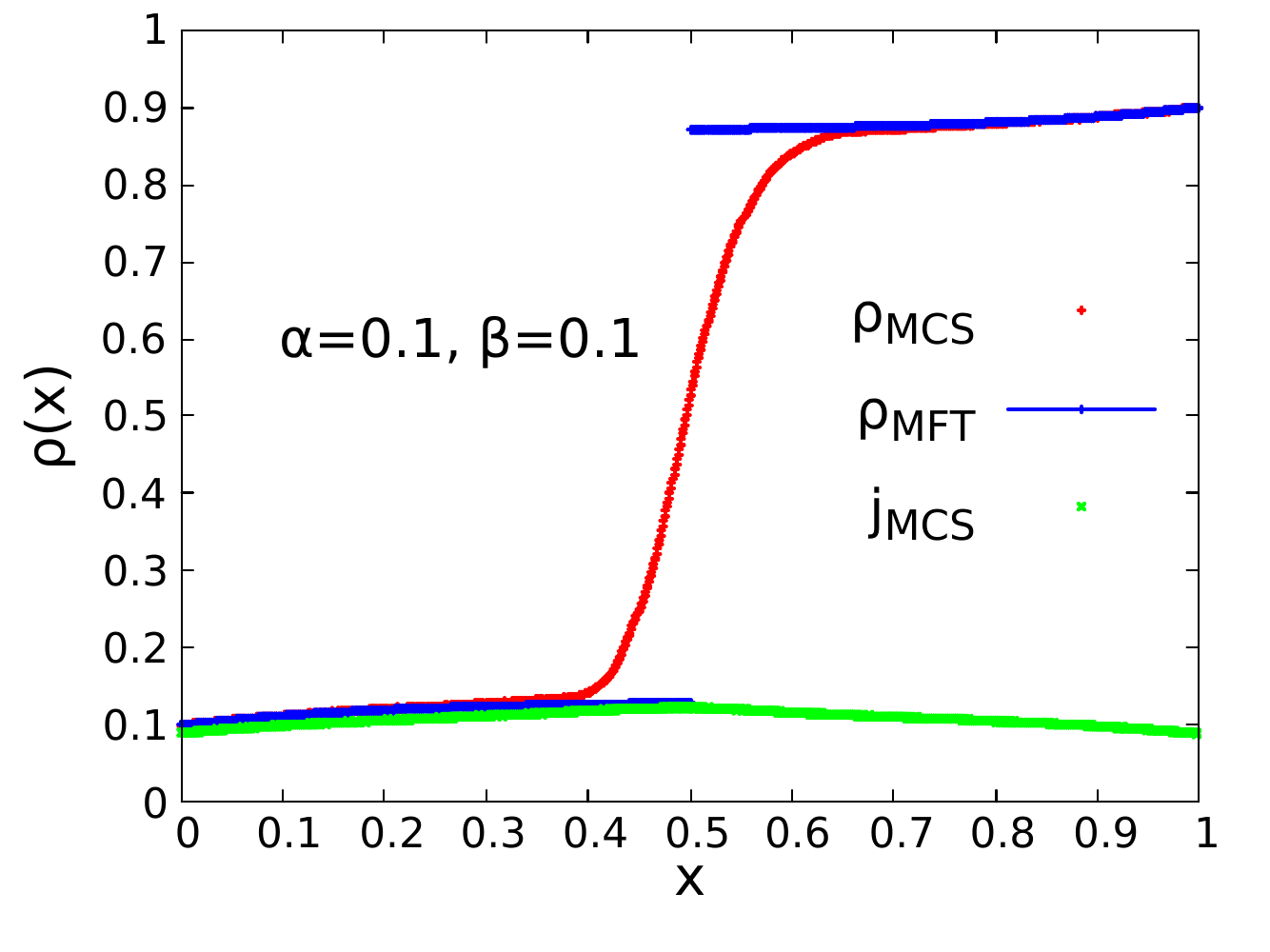}\\
 \includegraphics[width=0.85\columnwidth]{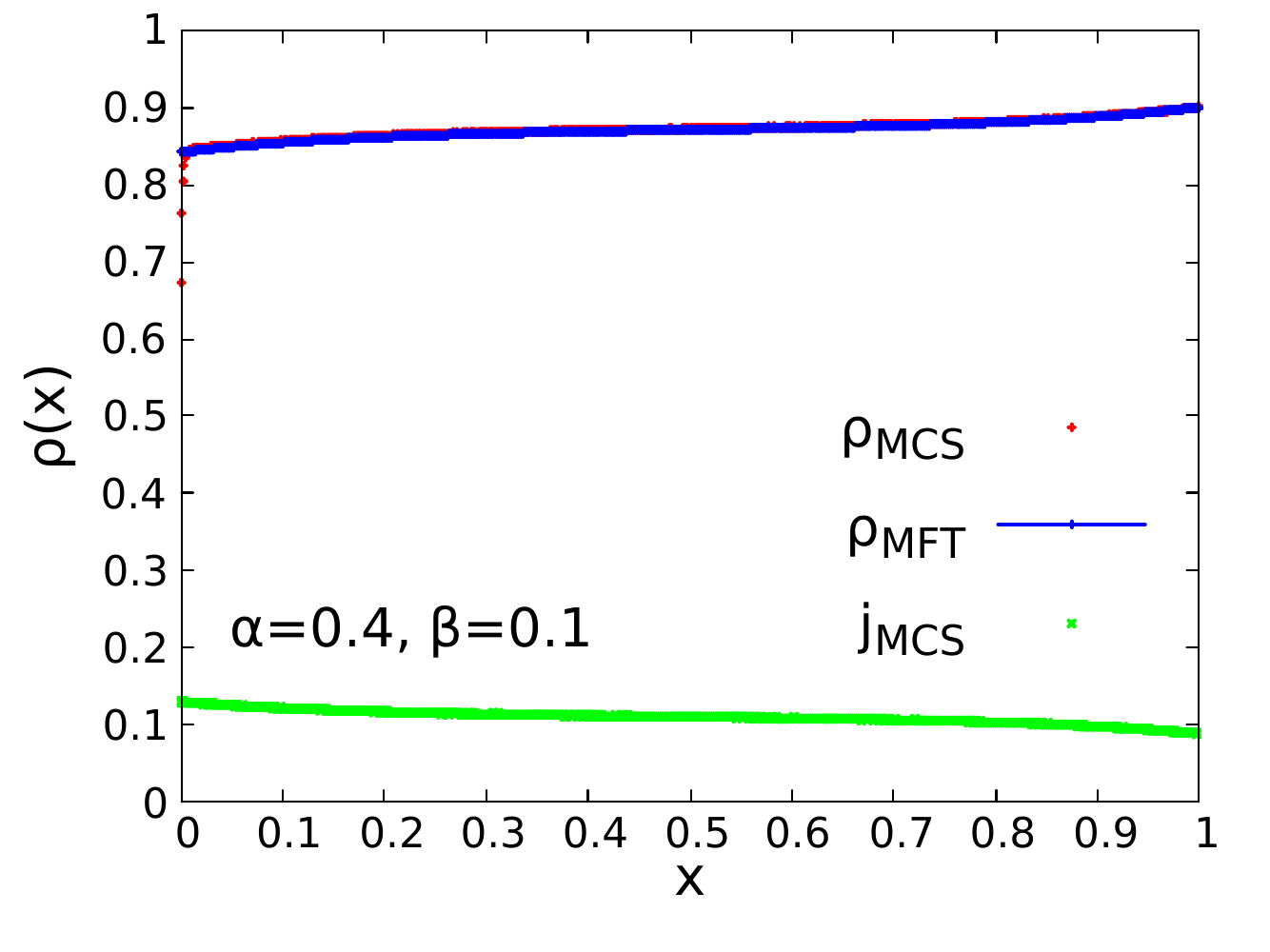}
 \caption{Steady state density $\rho(x)$ in the TASEP lane in the LD (top), LD-HD (middle) and HD (bottom) phases with $D=0.01,\,\Omega=0.3$. MFT (blue line) and MCS (red points) results are shown. For MFT solutions of TASEP, we have used $\rho(x)=\rho_\text{LD}(x)$ [Eq.~(\ref{rho-ld-smallD})] in the LD phase and $\rho(x)=\rho_\text{HD}(x)$ [Eq.~(\ref{rho-hd-smallD})] in the HD phase of the TASEP (see text). { The corresponding steady state TASEP current $j_\text{MCS}(x)$, as obtained from the MCS solutions, are shown in green.}}\label{d-01-tasep}
\end{figure}

\begin{figure}[htb]
 \includegraphics[width=0.85\columnwidth]{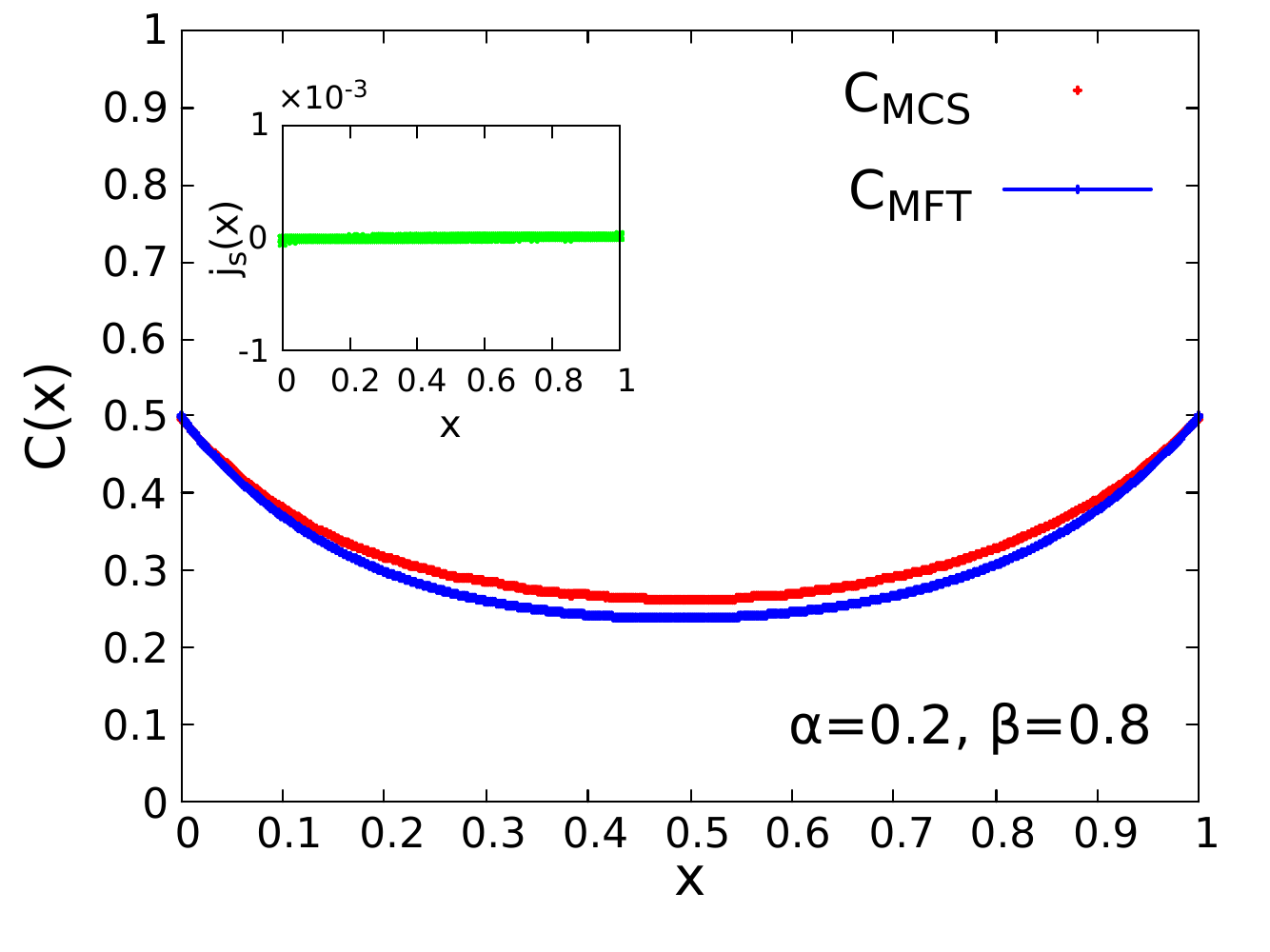}\\
 \includegraphics[width=0.85\columnwidth]{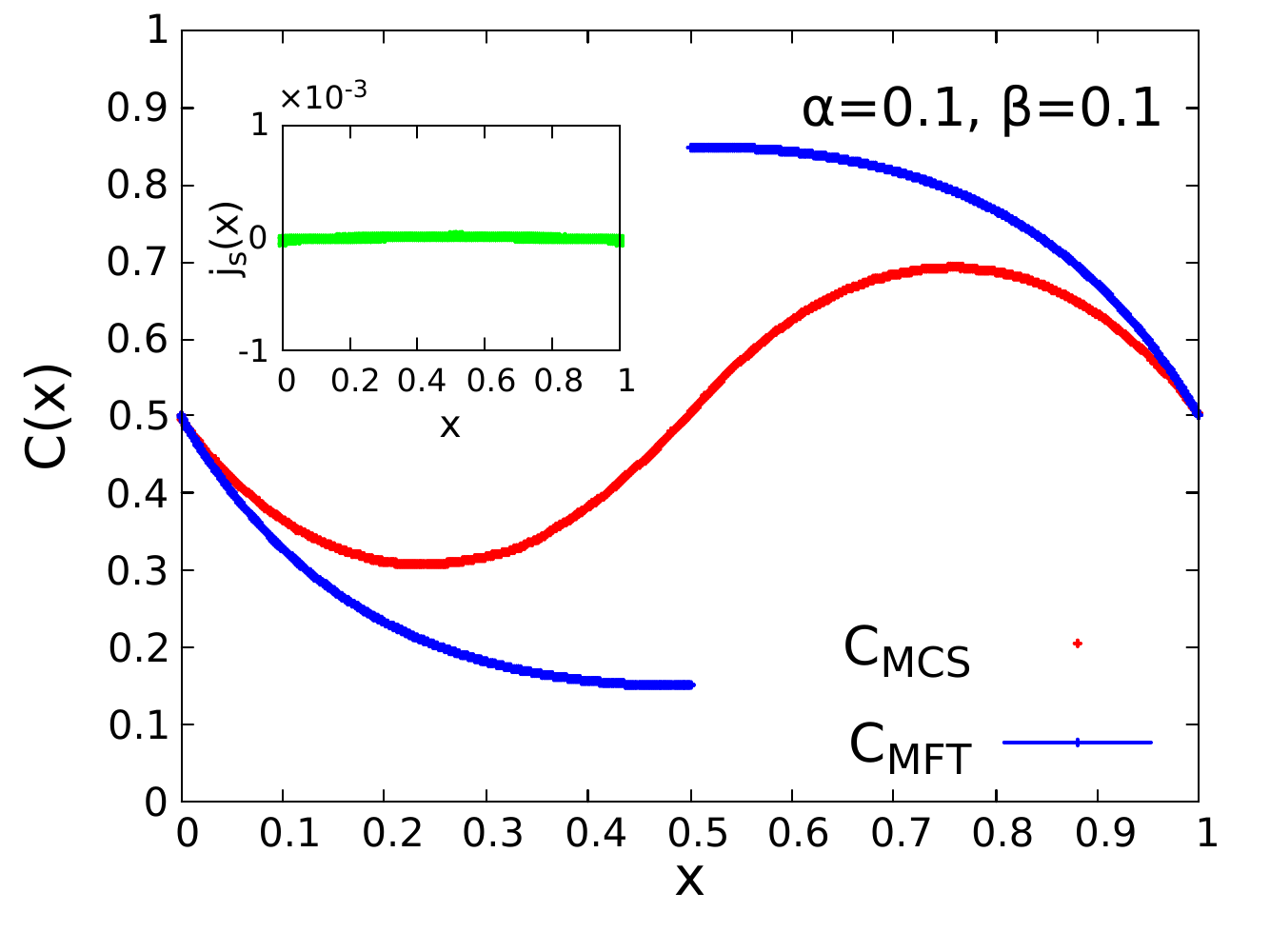}\\
 \includegraphics[width=0.85\columnwidth]{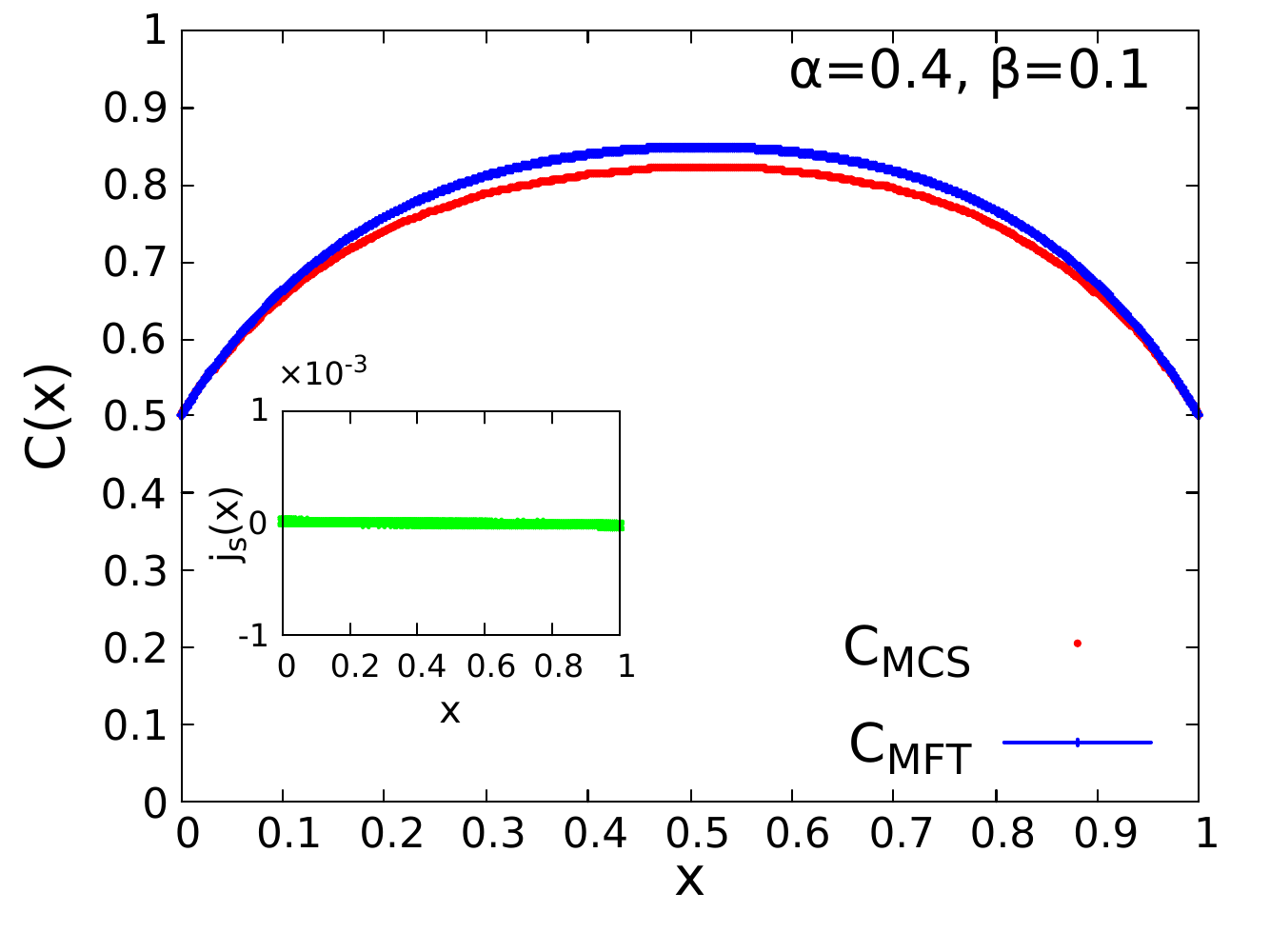}
 \caption{Steady state density $c(x)$ in the SEP lane, when the TASEP lane is in its LD (top), LD-HD (middle) and HD (bottom) phases with $D=0.01,\,\Omega=0.3$. MFT (blue line) and MCS (red points) results are shown. For MFT solutions of SEP, we have used $c(x)=c_-(x)<1/2$ and $c(x)=c_+(x)>1/2$ as defined in Eq.~(\ref{ceq-soln})]. { In the inset of each figure, the corresponding SEP current $j_s(x)$ is plotted as a function of $x$.} (see text).}\label{d-01-sep}
\end{figure}

{We again note a lower degree of quantitative agreement between the MFT and MCS solutions of $c(x)$, when the TASEP is in its LD-HD phase, relative to when it is in its LD or HD phases, in which case the agreement is better. As for our large-$D$ MFT solutions, we attribute this to the stronger space dependence of $c(x)$ near the location of an LDW in the TASEP, a feature not adequately captured by our MFT for $c(x)$. Nonetheless, the MFT and MCS solutions for $c(x)$ agree qualitatively.}

\subsection{Comparison of the MFTs}

When $D$ is neither too small or too large, neither of the approximations leading to the MFT solutions are expected to work well. Nonetheless, {\em both} MFTs should act as  guidelines to understand the MCS results. In Fig.~\ref{d1-phase-tasep}, we have shown the MFT phase diagrams for $D=0.05,\,\Omega=0.3$, obtained in the large $D$ approximation (solid red lines) and small $D$ approximation (broken black lines). While the two approximations clearly differ quantitatively, the topology of the TASEP phase diagrams remains the same, independent of the approximations used to obtain the MFT. This clearly lends credence to the physical pictures that emerge from this work.
\begin{figure}[htb]
 \includegraphics[width=0.9\columnwidth]{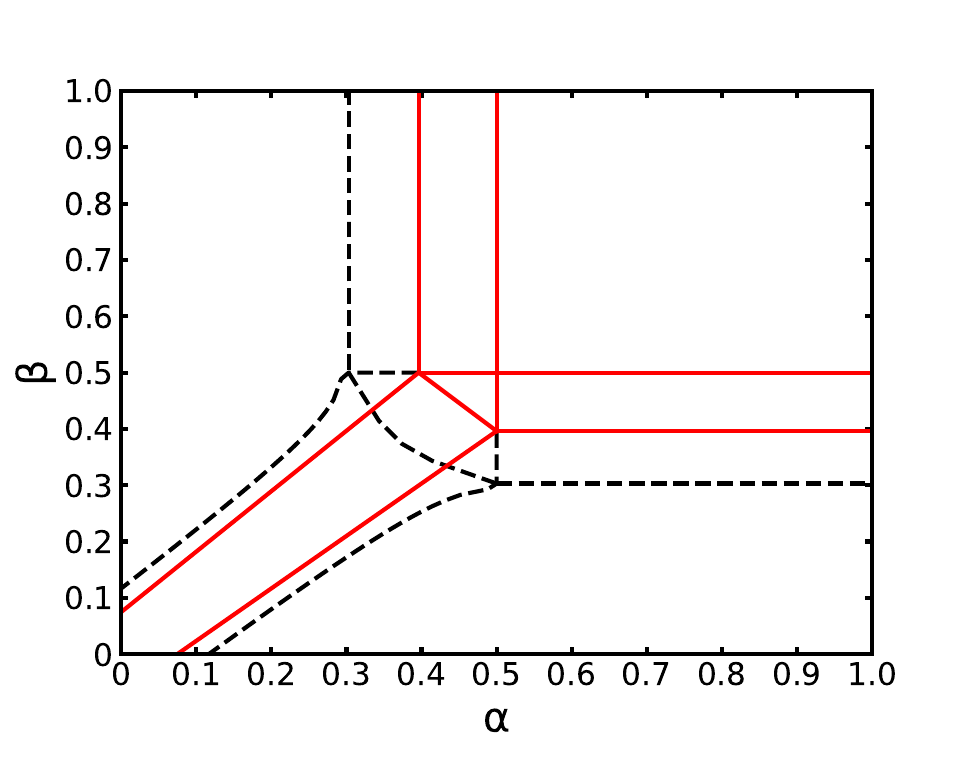}\\
 \caption{Mean-field TASEP phase diagrams with $D=0.05$, $\Omega=0.3$. MFT solutions for $\rho(x)$ with large $D$ approximation and small $D$ approximation are used to get the corresponding phase diagrams. Phase diagram with solid red (broken black) phase boundaries is obtained using large $D$ (small $D$) approximation (see text). }\label{d1-phase-tasep}
\end{figure}

We have further presented our results on $\rho(x)$ when the TASEP is in its LD and HD phases. Numerical results from our MCS simulations and the two MFT predictions (in the large-$D$ and small-$D$ approximations) are plotted together. We find that the MFT with large-$D$ approximation underestimates $\rho_\text{LD}(x)$ and overestimates $\rho_\text{HD}(x)$ with respect to the corresponding MCS results. The trend from the MFT with small-$D$ approximation is just the opposite. See Fig.~\ref{tasep-d-1} for plots of the MCS results on $\rho(x)$ together with the corresponding MFT predictions using MFTs with both large-$D$ and small-$D$ approximations. The results for the SEP density profiles $c(x)$ are shown in Fig.~\ref{sep-d-1}.

\begin{figure}[htb]
 \includegraphics[width=0.9\columnwidth]{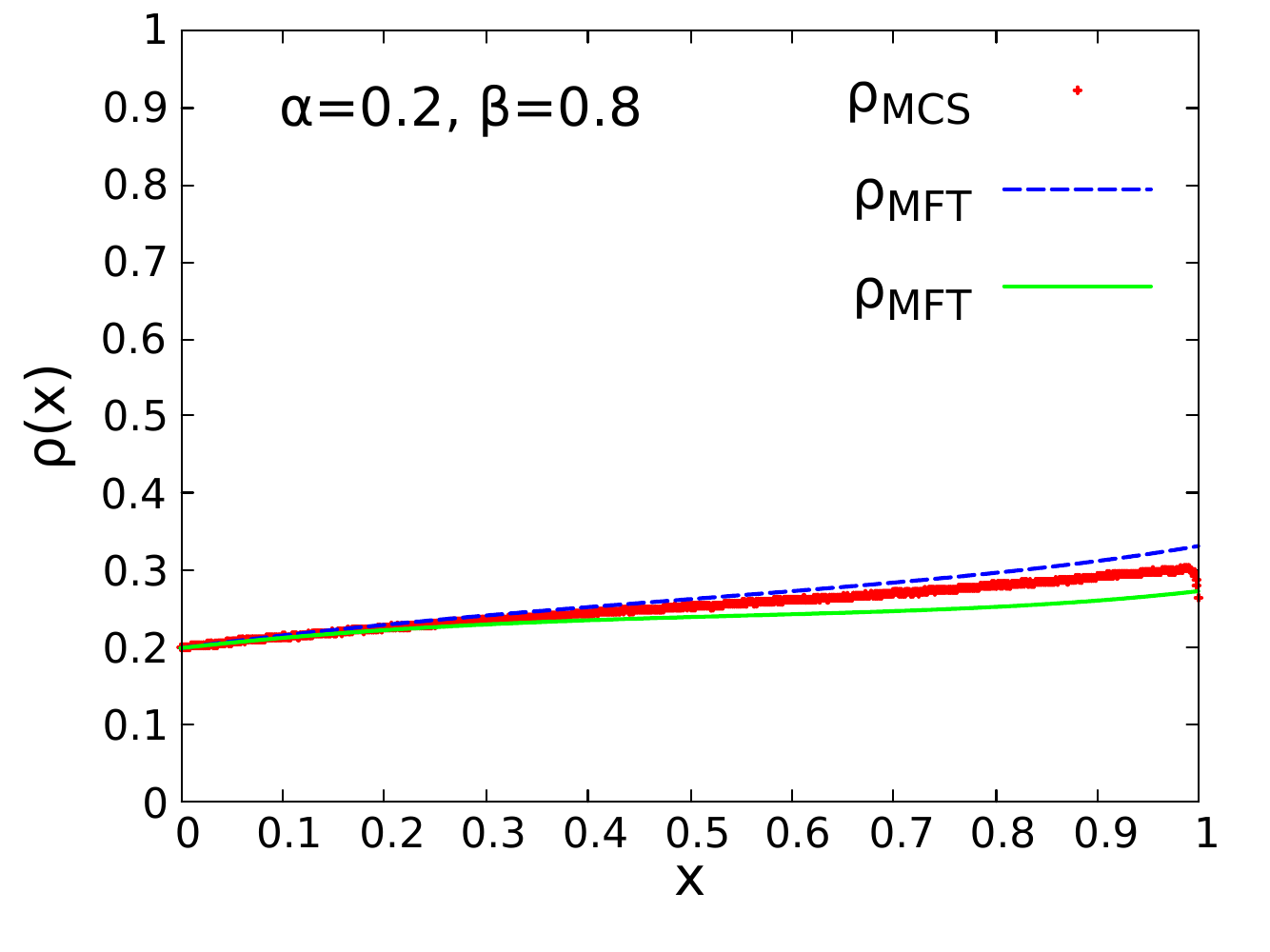}\\
 \includegraphics[width=0.9\columnwidth]{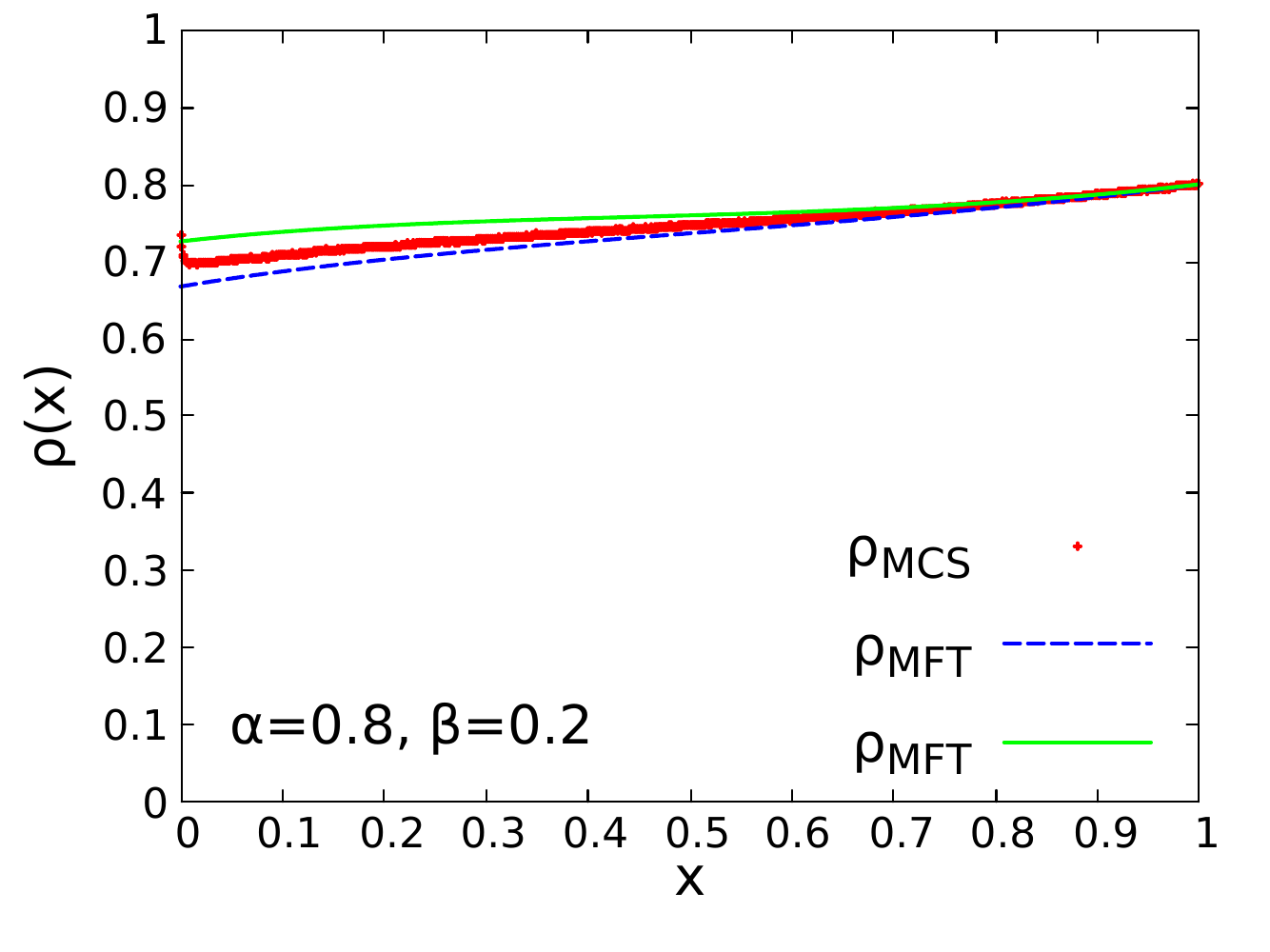}
 \caption{Plots of steady state density $\rho(x)$ versus $x$. (top) LD and (bottom) HD phases with $D=0.05$, $\Omega=0.3$. Broken blue lines represent $\rho_\text{MFT}$ obtained in the small-$D$ approximation, whereas the solid green lines represent $\rho_\text{MFT}$ obtained in the large-$D$ approximation; red points represent the corresponding MCS results.}\label{tasep-d-1}
\end{figure}

\begin{figure}[htb]
 \includegraphics[width=0.9\columnwidth]{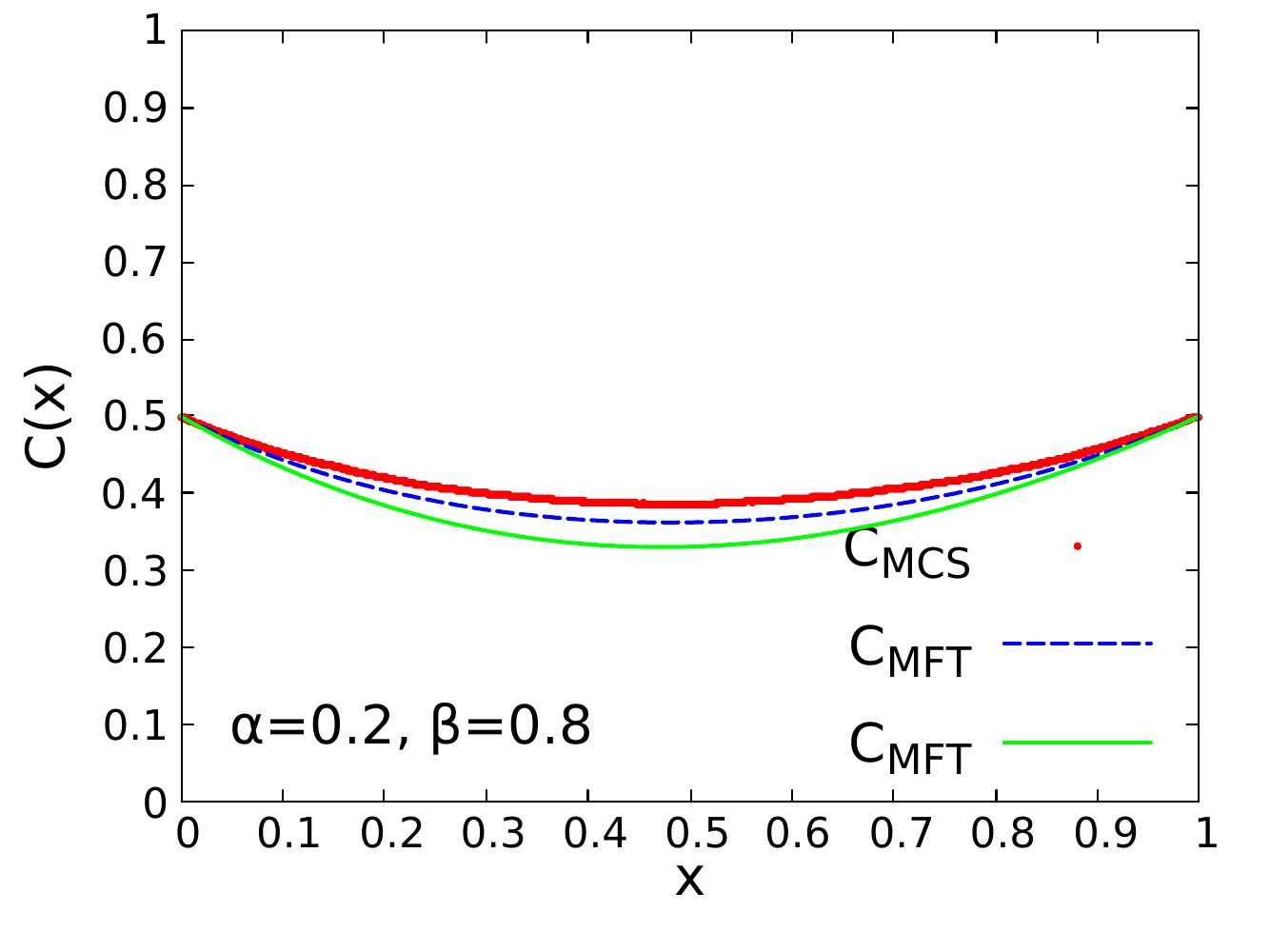}\\
 \includegraphics[width=0.9\columnwidth]{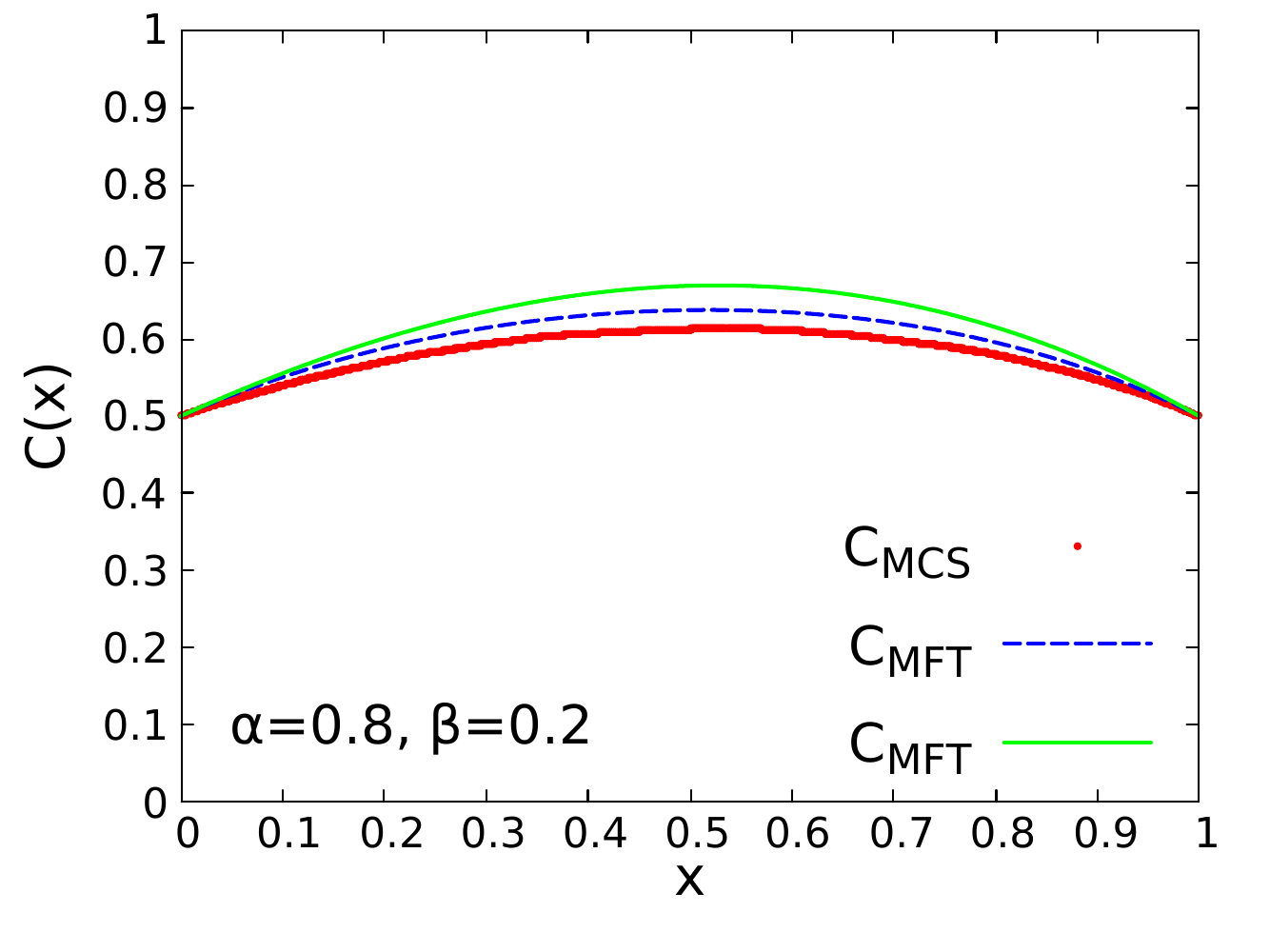}
 \caption{Plots of SEP steady state density $c(x)$ versus $x$, when the TASEP is in its (top) LD phase and (bottom) HD phase with $D=0.05$, $\Omega=0.3$. Broken blue lines represent $C_\text{MFT}$ obtained in the small-$D$ approximation, whereas the solid green lines represent $C_\text{MFT}$ obtained in the large-$D$ approximation; red points represent the corresponding MCS results. }\label{sep-d-1}
\end{figure}

{

\section{Three-phase coexistence and variation of $D$}

Direct visual comparison of the TASEP phase diagrams in Fig.~\ref{d10-phase-tasep} with $D=1$ and Fig.~\ref{d01-phase-tasep} with $D=0.01$ reveal that there is a substantial reduction in the phase-space region corresponding to the three-phase coexistence region in the $\alpha-\beta$ plane. This is unsurprising, as ultimately with $D\rightarrow 0$, our model reduces to an isolated open TASEP, which does not have any three-phase coexistence. Now consider starting from a high enough $D$ with a phase-space location characterised by particular values of $\alpha,\,\beta$ falling in the three-space region. As $D$ is reduced, the three-phase coexistence region shrinks and this particular point will now be outside the three-phase coexistence region. Indeed, we find that with $\alpha=0.44,\beta=0.46,\,\Omega=0.3$, we find  with $D=1$ a steady state TASEP density with a three-phase coexistence; see Fig.~\ref{3-phase}. In contrast, when $D$ is reduced to 0.0001, with the same $\alpha=0.44,\beta=0.46$, one finds a pure LD phase density, meaning the three-phase coexistence region has shrunk making $\alpha=0.44,\beta=0.46$ now falling in the LD phase region of the phase diagram with $D=0.0001$.
\begin{figure}[htb]
 \includegraphics[width=0.9\columnwidth]{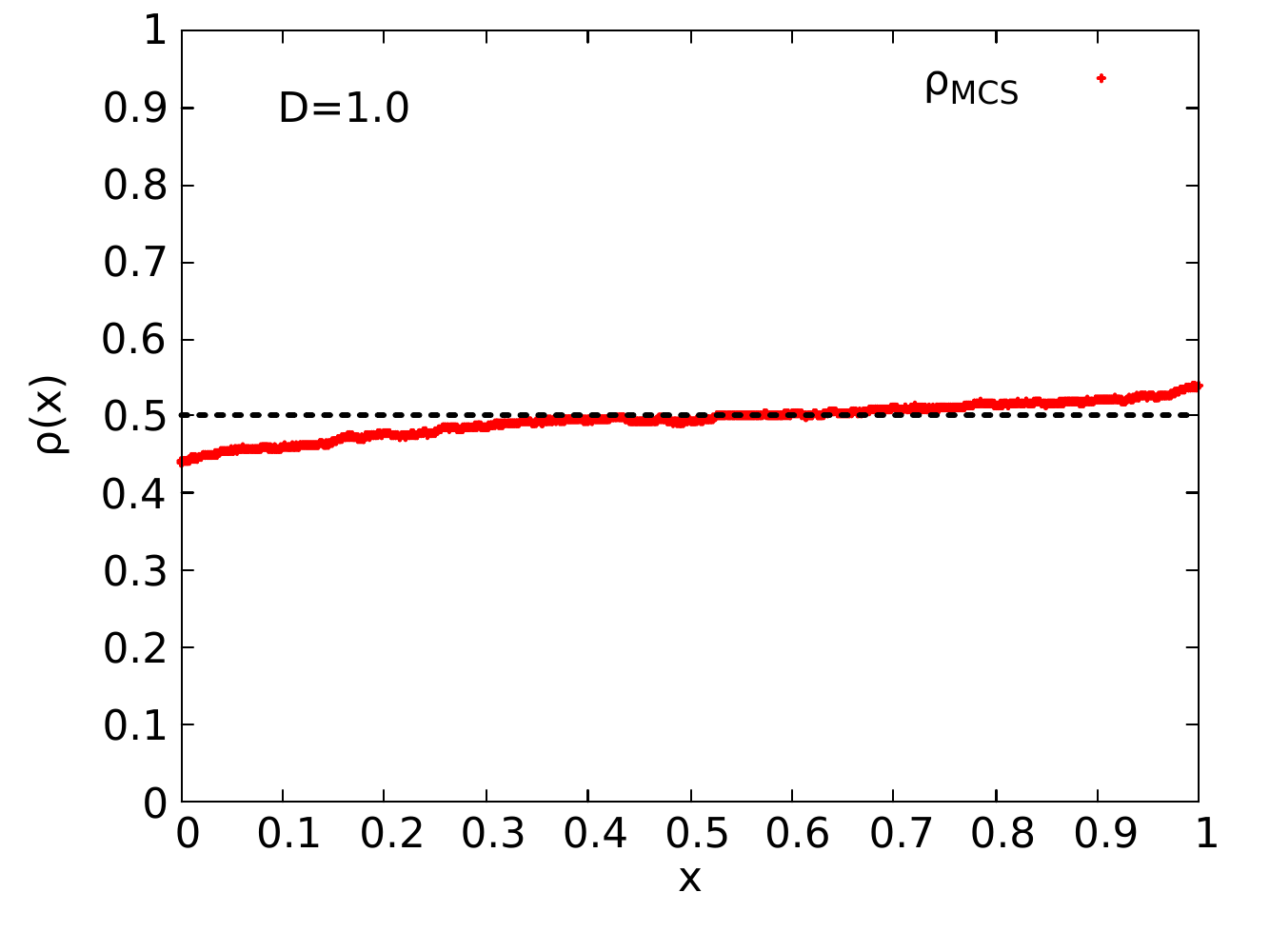}\\
 \includegraphics[width=0.9\columnwidth]{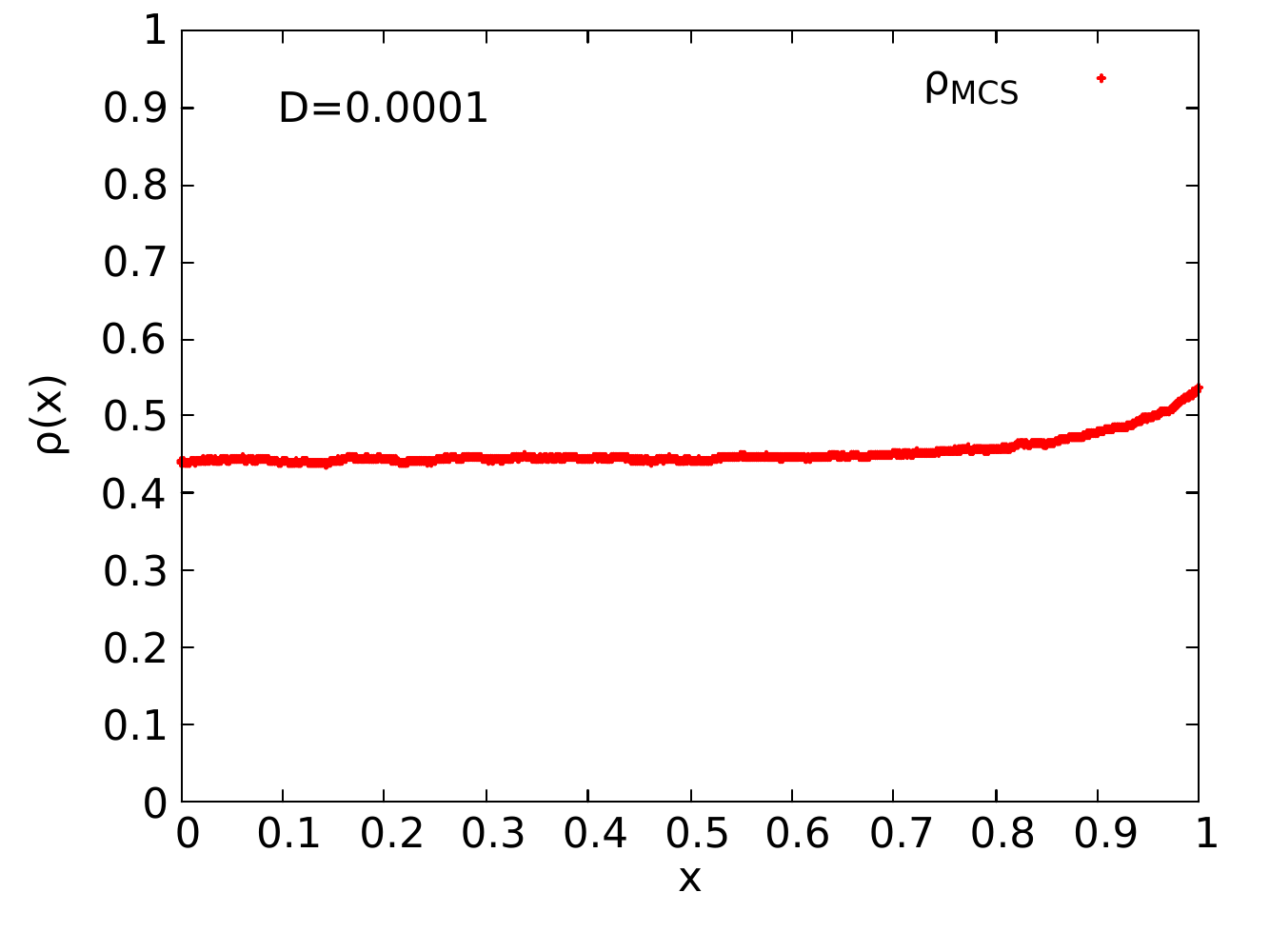}
 \caption{Changes in the TASEP density as $D$ changes ($\alpha=0.44,\beta=0.46,\,\Omega=0.3$). (top) three-phase coexistence with $D=0.1$, (bottom) LD phase density with $D=0.0001$.} \label{3-phase}
\end{figure}

\section{Phase diagrams in the $\Omega-D$ plane}

In the above, we have presented the phase diagrams in the $\alpha-\beta$ plane for fixed values of $\Omega$ and $D$, separately for small $D$ and large $D$.  It is also instructive to consider the phase diagrams in the $\Omega-D$ plane for fixed values of $\alpha$ and $\beta$; see the phase diagrams in Fig.~\ref{om-D} for two sets of representative values of $\alpha$ and $\beta$. 
As before, the phase boundaries may be obtained by equating the steady state currents of the different phases as functions of $\Omega,\,D$ and parametrised by $\alpha,\,\beta$.
In these phase diagrams, we notice that mixed phases with an MC part appear for larger $\Omega$. This is not surprising, as larger values of $\Omega$ necessarily means more particle exchanges between the SEP and TASEP channels. This in turn tends to bring $\rho(x)$ and $c(x)$ closer, increasing the possibility of the MC phase and hence mixed phases with an MC part.}  
\begin{figure}[htb]
\includegraphics[width=0.9\columnwidth]{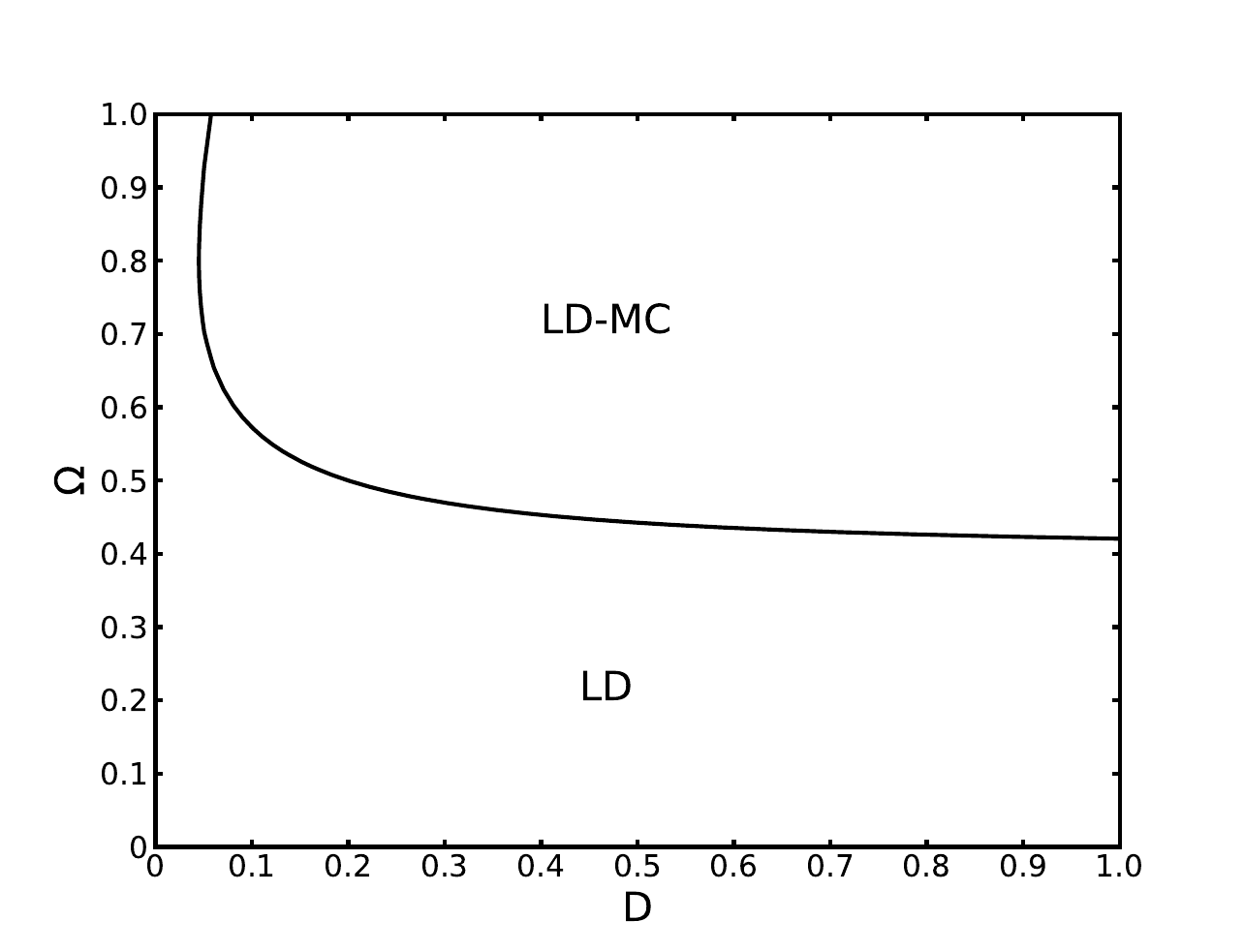}\\
 \includegraphics[width=0.9\columnwidth]{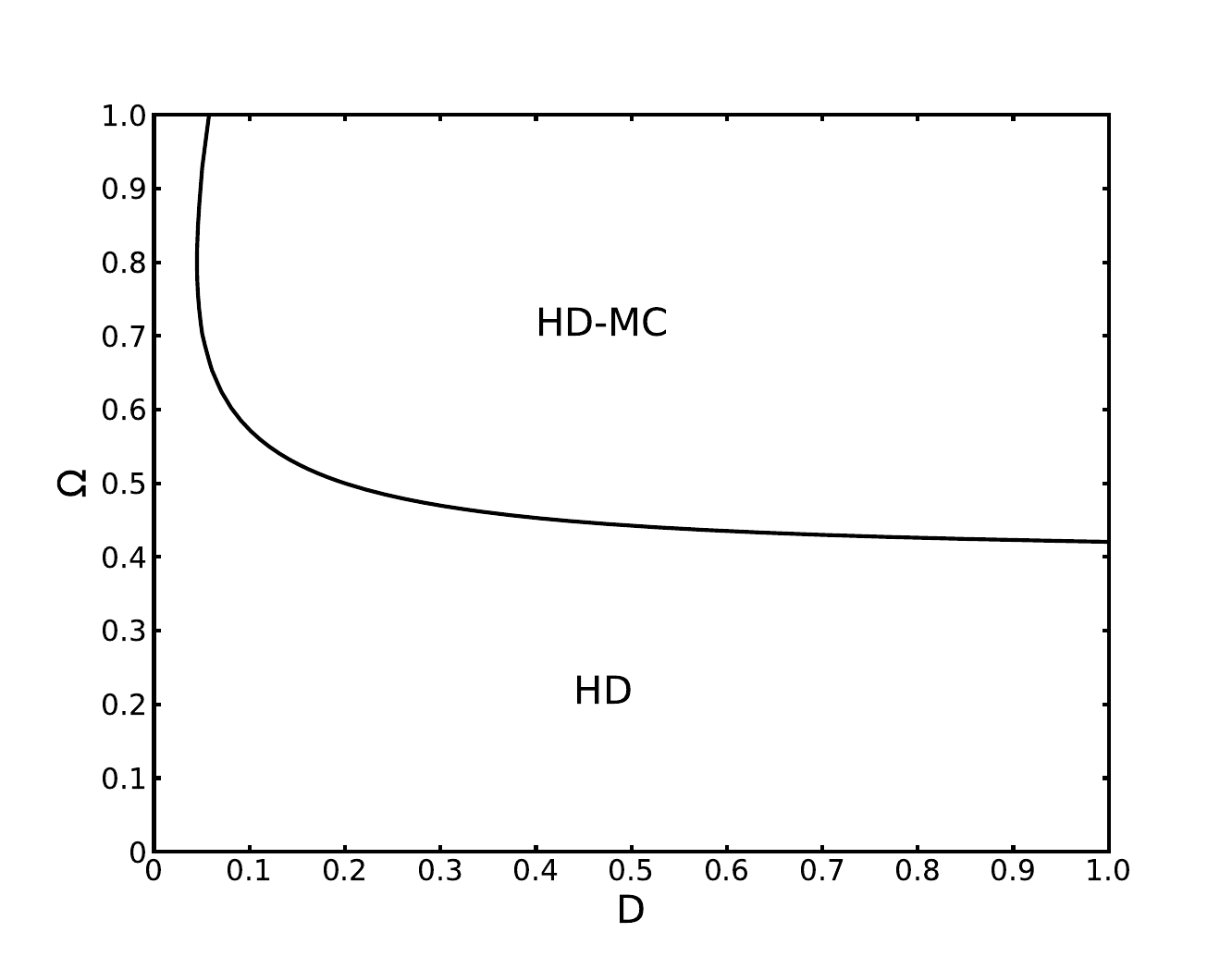}
 \caption{Phase diagrams in the $\Omega-D$ plane for fixed $\alpha,\,\beta$: (top) $\alpha=0.3, \beta=0.9$, (bottom)$\alpha=0.9, \beta=0.3$. }\label{om-D} 
\end{figure}

\section{Delocalisation of domain walls for $D\rightarrow 0$}\label{ddw}

{We have found that in the limit of $D\rightarrow \infty$ this model reduces to the LK-TASEP model~\cite{frey-lktasep}, whereas in the limit $D\rightarrow 0$ it reduces to an isolated open TASEP. A smooth crossover from LK-TASEP behaviour to an open TASEP is expected as $D$ is reduced. This can be seen from the phase diagrams given above for various values of $D$. As $D$ is reduced the regions for two- and three-phase coexistence regions shrink, which are expected to vanish for $D\rightarrow 0$, i.e., in the limit of a pure TASEP. Indeed, the two-phase coexistence region should shrink to the limit $\alpha=\beta$ and the three-phase coexistence to a point $\alpha=\beta=1/2$ in the isolated, open TASEP limit with $D\rightarrow 0$. Our MFT is consistent with these physically motivated expectations. In Fig.~\ref{comp-phase}, mean-field TASEP phase diagrams for various values of $D$ ranging from $0.1$ to  $0.000001$ are shown. These phase diagrams are drawn with the MFT valid for small $D$ (or, equivalently, $\Omega/D\gg 1$). It is evident that as $D$ is progressively reduced, the  two- and three-phase coexistence regions increasingly shrink, eventually practically vanishing for $D=0.000001$, for which the resulting phase diagram is virtually indistinguishable from that of an isolated open TASEP.

\begin{figure}[htb]
 \includegraphics[width=0.9\columnwidth]{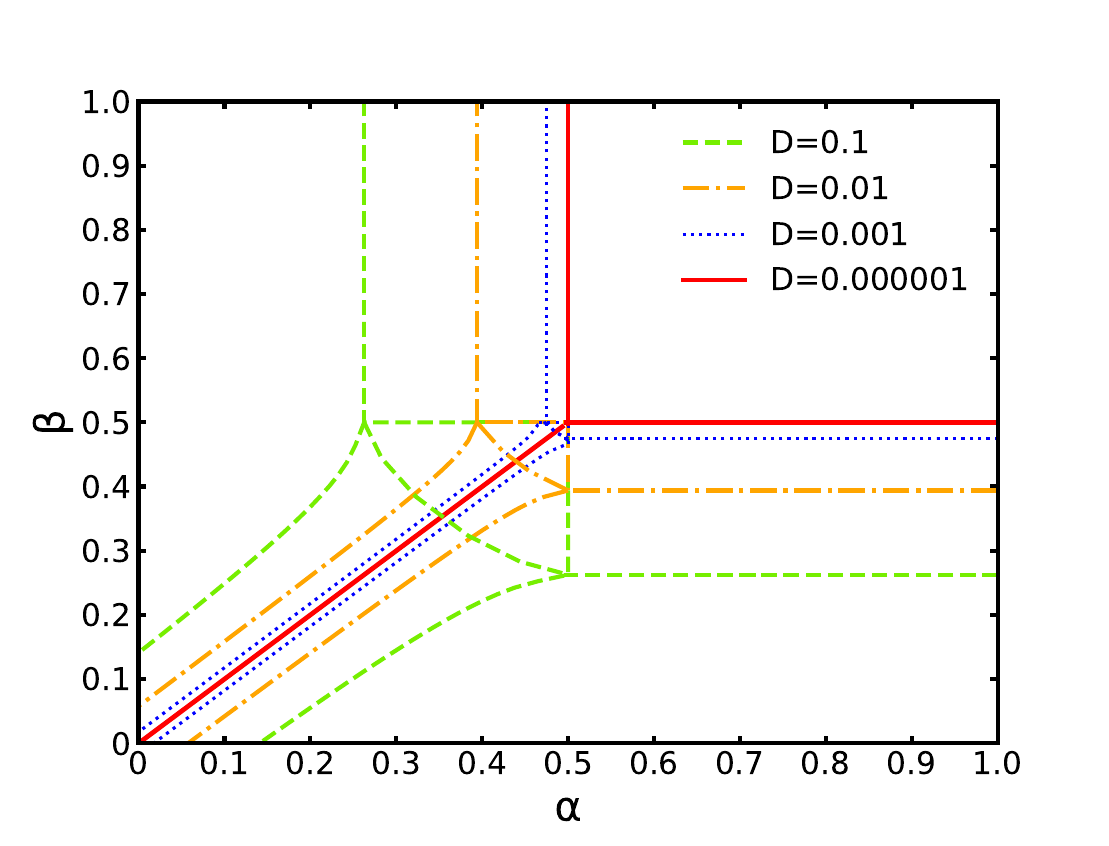}
 \caption{Mean-field TASEP phase diagrams in the $\alpha-\beta$ plane for various values of $D$ ranging from 0.1 to  0.000001 are shown. These phase diagrams are drawn with the MFT valid for small $D$ (or, equivalently, $\Omega/D\gg 1$).
 Clearly for $D=0.000001$, the phase diagram is virtually indistinguishable from its counterpart for an isolated open TASEP.}\label{comp-phase}
\end{figure}

The TASEP density profiles in the two-phase coexistence region for any $D>0$ (including the LK-TASEP limit of $D\rightarrow \infty$) is a pinned or static domain wall, i.e., an LDW. This pinning of the domain walls is attributed to spatially nonuniform TASEP densities in the steady states. However, in the limit $D\rightarrow 0$, it must  be a DDW, existing on the line $\alpha=\beta$ for $0<\alpha=\beta<1/2$, as it is for an isolated open TASEP. While a fully delocalised domain wall is possible only for $D\rightarrow 0$, we observe signatures of gradual delocalisation as $D$ is reduced. To see this, we obtain the TASEP density profiles on the line $\alpha=\beta=0.1$ with $\Omega=0.3$ for system size $L=1000$ and various values of $D$.  We find that as $D$ is reduced, the long-time averaged profile of the domain wall becomes an increasingly inclined line, signifying larger fluctuations in its position. { We visually measure the extent of domain wall position fluctuations or the ``width'' $W$, which is roughly the projection of the inclined line of the domain wall on the $x$-axis (see the inset in Fig.~\ref{dws}), and plot them against $D$. See Fig.~\ref{dws} for plots of the domain walls with $D=1,0.005, 0.001$, and Fig~\ref{dw-position} for a semilog plot of $W$ versus $D$. While our study is purely phenomenological, it does indicate increasing delocalisation as $D$ is reduced. Note that this effect cannot be captured within MFT, as MFT by construction neglects all fluctuations. The approaches developed in Ref.~\cite{frey-2lane1} to study fluctuations systematically going beyond MFT may be helpful in this study. } }

\begin{figure}[]
 \includegraphics[width=0.9\columnwidth]{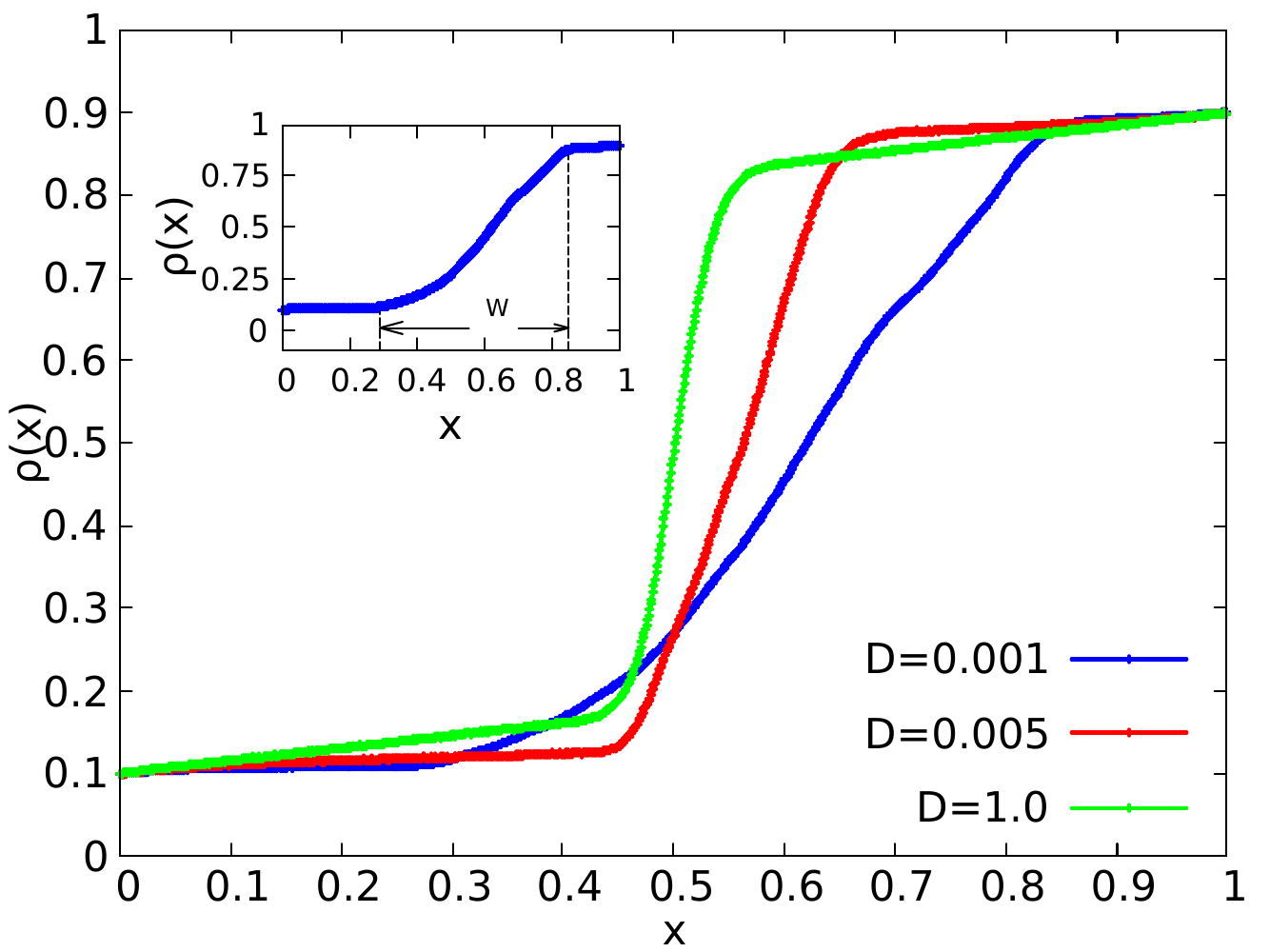}
 \caption{Gradual delocalisation of the domain wall due to increasing fluctuations as $D$ is reduced for a fixed system size $L=1000$ and $\alpha=\beta=0.1$. MCS results for $D=1.0,0.005, 0.001$ are shown. (Inset) A pictorial definition of the DW width $W$ (for $D=0.001$) is shown. Clearly, as $D$ is reduced, $W$ is  increased for a fixed system size $L$, indicating gradual delocalisation of the domain wall.}\label{dws}
\end{figure}


\begin{figure}[htb]
 \includegraphics[width=0.9\columnwidth]{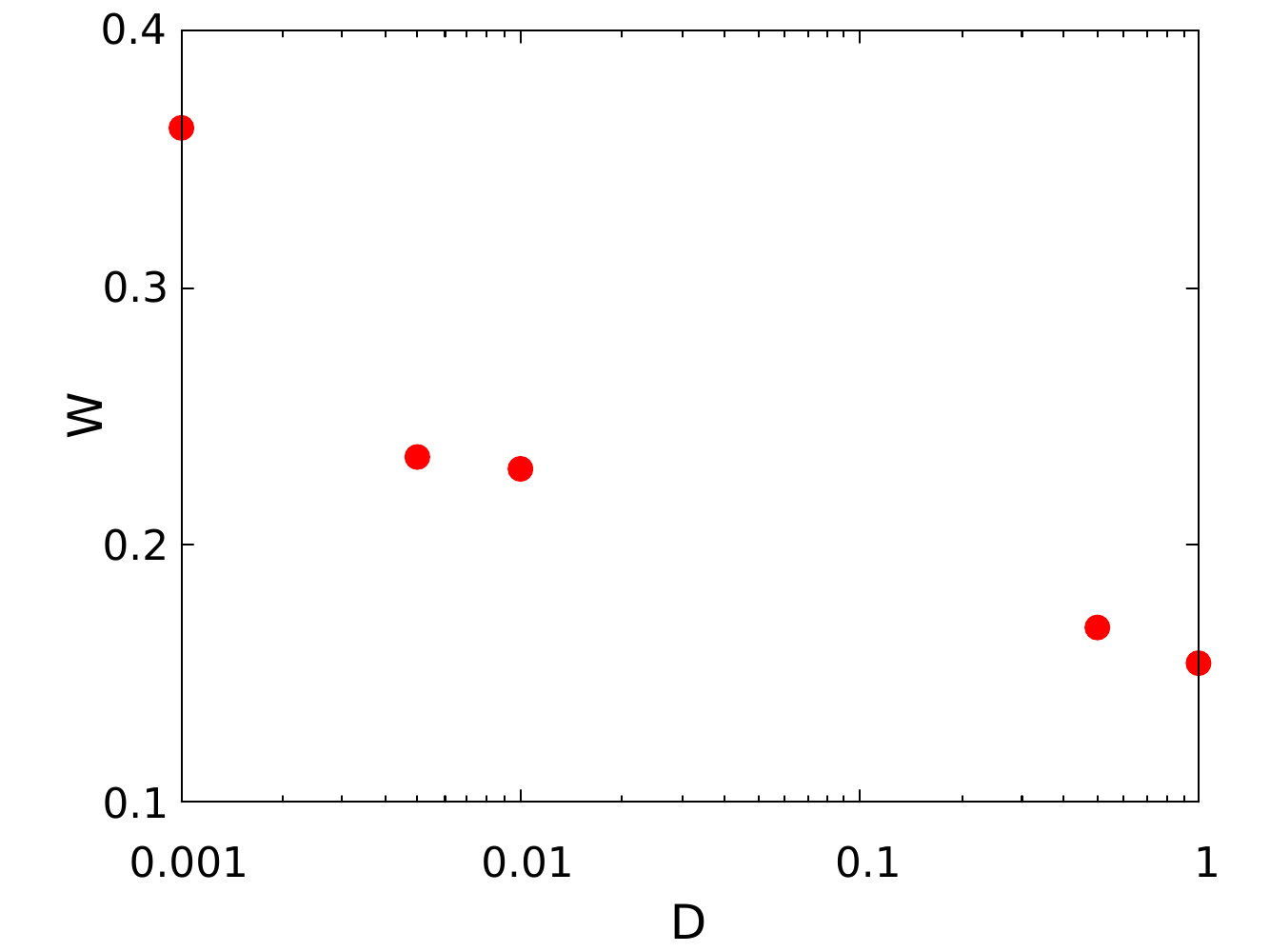}
 \caption{Semilog plot of $W$ as a function of $D$ for $\Omega=0.3,\alpha= 0.1 =\beta$. Clearly, $W$ rises as $D$ is reduced, indicating gradual delocalisation of the domain wall (see text).}\label{dw-position}
\end{figure}

{ 
\section{Effects of a biased SEP}

In our above studies, we have confined ourselves to an open {\em unbiased} SEP, i.e., the SEP carries no net steady current when isolated. In a general situation, however, the SEP may be {\em biased} due to different incoming and outgoing fluxes, which then has a net steady state current.  This can be achieved  by setting unequal entry and exit rates at the two ends of the SEP channel, resulting into unequal densities at $x=0$ and $x=1$, i.e., $c(0)\neq c(1)$. This gives rise to a spatially nonuniform SEP density that depends linearly on $x$: 
\begin{equation}
c(x)=c(0)+ [c(1)-c(0)]x=c(0)+{\cal A}x, \label{sep-den-bias}
\end{equation}
where ${\cal A}=c(1)-c(0)$, the slope, is a measure of the bias.
This, when coupled with the TASEP via particle exchanges, can potentially affect the TASEP steady state densities, which we briefly discuss below. First of all, with a biased SEP, $\rho(x)=1/2=c(x)$ are no longer steady state solutions. For simplicity, we now consider the limit $D\rightarrow \infty$. Like the unbiased case, in this limit $c(x)$ is independent of $\rho(x)$, and is given by (\ref{sep-den-bias}).  The MFT equation (\ref{rho-mft}) for $\rho(x)$ now reads (with a constant $\Omega$)
\begin{equation}
 \Omega [c(0)+ {\cal A}x - \rho(x)]+(2\rho-1)\partial_x\rho=0.
\end{equation}
Evidently, $\rho=1/2$ is {\em not}  a solution for $\rho(x)$, as already mentioned earlier.
Assuming a small ``bias'' ${\cal A}$, for which $c(0)\approx 1/2$, we attempt to solve $\rho(x)$ approximately by perturbatively expanding about the solution of $\rho(x)$ for an unbiased SEP, i.e., $\rho(x)=\rho_\text{LK}(x)$. We write 
\begin{equation}
 \rho(x)=\rho_\text{LK}(x) + \phi(x).
\end{equation}
In the limit of $ {\cal A}\rightarrow 0$ and $c(0)\rightarrow 1/2$, $\phi(x)$ should vanish with $\rho (x)=\rho_\text{LK}(x)$ along with $D\rightarrow \infty$. For small $\cal A$, $\phi(x)$ should be a small correction over $\rho_\text{LK}(x)$. 
The equation that $\phi(x)$ satisfies should clearly depend upon the form of $\rho_\text{LK}(x)$. If  $\rho_\text{LK}(x)$ corresponds to the LD or HD phase solutions in the TASEP-LK model~\cite{frey-lktasep}, then $\phi$ satisfies
\begin{equation}
 \partial_x \phi(x)= \frac{f(x)}{2\rho_\text{LK}(x) -1},\label{biased-sep}
\end{equation}
to the linear order in $\phi(x)$,
where $f(x)=-\Omega(c(0)+ {\cal A}x - 1/2)$. Equation~(\ref{biased-sep}) can be solved together with the boundary conditions $\phi(0)=0$ or $\phi(1)=0$, depending upon $\rho_\text{LK}(x)$ being the LD or HD phase solution, respectively. In the special case of $\rho_\text{LK}(x)=1/2$ corresponding to the occurrence of the MC phase solution in bulk of the TASEP-LK model, $\phi(x)$ is given by
\begin{equation}
 \phi(x)=c(0)+ {\cal A}x - \frac{1}{2},\label{biased-sep-mc}
\end{equation}
in the bulk of the TASEP, neglecting higher order terms in $\phi$. 
From (\ref{biased-sep}) or (\ref{biased-sep-mc}), we make the following observations: 
The solution of $\phi(x)$ depends explicitly on the sign of the slope $\cal A$, which is also the SEP current. This means the TASEP density $\rho(x)$ should depend on whether the SEP current is parallel or antiparallel to the TASEP current. Furthermore, from (\ref{biased-sep}) it is clear that $\phi(x)$ - and hence the TASEP density $\rho(x)$ -  should have generic nonlinear dependence on $x$. This means the phase boundaries will likely have more complex structures. These comments hold for small $\cal A$, meaning small SEP biases. For finite biases, one may have to recourse to numerical solutions of MFT equations.  A more complete analysis of the TASEP density profiles as functions of the SEP bias including large biases will be presented elsewhere.

}

\section{Summary and outlook}\label{summ}

In summary, we have proposed and analysed an open one-dimensional system with two lanes, one modeling a 
one-dimensional lattice executing TASEP dynamics and the other with diffusive or SEP kinetics, representing a reservoir, which are coupled by exchange of
particles subject to exclusion. This diffusion is unbiased, that is a particle can hop to its right or left with equal probability, subject to exclusion. We show that the ratio of the effective exchange rate $\Omega$ to the diffusion coefficient $D$, or for a fixed $\Omega$, $D$ itself appears as
the tuning parameter by varying which our system goes from a diffusion
dominated behaviour for large $D$ to TASEP dominated behaviour in the limit of small $D$. We show that for a fixed non-zero $\Omega$, with $D\rightarrow0$
the SEP density is slaved to the TASEP density and the resulting
steady-state phase diagram of the TASEP lane is same as that of an isolated open TASEP. In the opposite extreme
limit of fast diffusion, i.e., $D \rightarrow \infty$, the density profile of the diffusive lane is spatially constant with a value 1/2,
whereas that in the  driven lane is identical to that of a TASEP with Langmuir Kinetics in the bulk.
For intermediate values of $D$, our model has nonuniform density profiles
in both the TASEP and the SEP  in the steady states with rather complex position-dependence. These nontrivial space dependences are entirely due to the coupling between the
SEP and TASEP kinetics, for without the coupling, both SEP and TASEP generally yield
flat density profiles (except for the delocalised domain wall in an open TASEP). For intermediate values of $D$, the MFT equations cannot be solved exactly. This has led us to solve the MFT equations for small and large $\Omega/D$ separately, and obtain two sets of solutions with one of them giving modifications of the  TASEP and SEP densities for small $\Omega/D$ and the other for large $\Omega/D$ in perturbative approaches. We find that the MFT solutions agree reasonably with the MCS studies results for small and large $\Omega/D$. However, unsurprisingly when $\Omega/D$ is intermediate none of the solutions agree quantitatively with the numerical results. We have also numerically explored how a domain wall in the TASEP lane, which is obviously localised for any finite $D$, but must be fully delocalised when $D\to 0$, actually gradually delocalises as $D$ is reduced. Such an effect cannot be studied within the MFT, as it neglects all fluctuations. It would be interesting to  theoretically  study this delocalisation by going beyond MFT descriptions and considering fluctuations. We have further discussed phase diagrams of the SEP in the plane of the control parameters of the TASEP, {\em viz.}, $\alpha,\beta$. We have argued that the phase diagram of the SEP is identical for any finite $D$ (including the limiting case of $D\rightarrow 0$): it has just three phases, {\em viz.} deficit, excess and neutral particle phases (measured with respect to the mean SEP density $\overline c=1/2$ in an isolated SEP with unbiased entry and exit rates). We have shown that these phases have a direct correspondence with the phases of the TASEP, a principal result from this work. 

The take home message from our studies here is that the mutual coupling between driven and diffusive dynamics can be tuned to make not only the TASEP lane to pick up non-trivial position dependence in its steady state densities, the diffusive lane can also maintain nonuniform steady states. This means a reservoir, which is modeled by a SEP here, can sustain spatial gradients in its densities, when it exchanges particles with a driven channel in the bulk. An interesting future study would be to impose  particle number conservation at a global level, i.e., in the combined system of a TASEP and a SEP. This would be an example of a system with finite resources having an internal dynamics in the reservoir~\cite{parna,astik-parna}. 

{ Our study can be extended in various ways. For instance, one can consider the effects of ``shaking'' the SEP channel~\cite{tirtha-maes},  that is known to strongly affect the SEP density. Similarly, shaking an isolated open TASEP leads to a bulk transition to the MC phase~\cite{tirtha-maes}. It is of interest to find out how such shaking can affect the coupled SEP-TASEP model, and alter the results obtained by us.} Yet another direction of future research could be effects of space-dependent hopping in the SEP or the TASEP channels~\cite{sudip-jstat}. 

We have restricted ourselves to studying a 1D model here. As we mentioned earlier, our model, in addition to its importance as a 
1D nonequilibrium model with coupled driven and diffusive dynamics, also serves as a minimal
model for a molecular motors - microtubules assembly inside eukaryotic cells. The SEP here models the diffusion
of the molecular motors in the bulk, whereas the TASEP represents unidirectional
driven motion. Since we have modeled
the (three-dimensional) reservoir by SEP, a 1D model, it raises an important phenomenological point: there are significant dynamical differences
between the two due to single file diffusion in 1D, which gives the mean square displacement $ \Delta \propto \sqrt
t$ (time) \cite{1ddiff} in unbiased diffusion. This questions applicability of our results for a realistic three-dimensional system. Nonetheless, it is known that for an infinitesimal bias $\Delta \propto t$ is restored
\cite{bias1,bias2}, whereas for an infinitesimal bias, our results on the steady states of the SEP should be practically same as here. This clearly allows using our 1D results to draw physical insight about the corresponding three-dimensional situations. Nonetheless, it would definitely be interesting, from both theoretical as well as phenomenological standpoints, to extend and study our model to higher dimensions. This should give a better handle to modeling intra-cellular transport more realistically. In this study, we have considered an unbiased SEP, i.e., a SEP with equal entry and exit rates. In a more general situation, the entry and exit rates could be different, which can result in a biased SEP, which has an inclined line-shaped density profile. It would be interesting to couple such a biased SEP with a TASEP via lane exchanges and investigate the resulting steady states. Our study here is restricted to equal-sized TASEP and SEP lanes. Introduction of unequal lengths is expected to give additional complex behaviour. We hope our studies here will provide impetus to address these questions theoretically in the future.

\section{Acknowledgement}

S.M. thanks SERB (DST), India for partial financial support through the CRG scheme [file: CRG/2021/001875].

\appendix

\section{Space-dependent exchange}

We now consider the effects of side-dependent exchange rates $\omega_i$. {We continue to assume equal attachment and detachment rates. As before, we use scaled attachment-detachment rates defined by $\omega_i=\Omega_i/L^2$ and TASEP hopping rate $1/L$ together with $\alpha/L,\,\beta/L$ as the entry and exit rates in TASEP to ensure competition with the diffusion in SEP}. This results into the MF equations (\ref{rho-mft}) and (\ref{c-mft}). We further consider the asymptotic limit of fast diffusion given by $D\rightarrow \infty$. In that limit, the SEP density is independent of the TASEP density with $c(x)=1/2$ everywhere, independent of the TASEP density. In that limit, using $c(x)=1/2$, Eq.~(\ref{rho-mft}) reduces to
\begin{equation}
 (1-2\rho)\left[\frac{\partial\rho}{\partial x}- \frac{\Omega(x)}{2}\right]=0. \label{appen-rho-mft}
\end{equation}
Equation (\ref{appen-rho-mft}) has two solutions: $\rho(x)=1/2$, which gives the MC phase here, and $\rho(x)=\int dx\, \Omega(x)/2 + \tilde C$, where $\tilde C$ is a constant of integration. By using the boundary conditions $\rho(0)=\alpha$ or $\rho(1)=1-\beta$, we can evaluate $\tilde C$: Using $\rho(0)=\alpha$
\begin{equation}
 \tilde C\equiv \tilde C_\alpha = \alpha - \left[\int dx \,\Omega(x)/2\right]_{x=0}.\label{const-0}
\end{equation}
Similarly, using $\rho(1)=1-\beta$, we get
\begin{equation}
 \tilde C\equiv \tilde C_\beta = 1-\beta - \left[\int dx \,\Omega(x)/2\right]_{x=1}.\label{const-1}
\end{equation}
Thus using (\ref{const-0}) and (\ref{const-1}) we get two solutions
\begin{eqnarray}
 \rho_\alpha(x)&=&\int dx \,\Omega(x)/2+\tilde C_\alpha,\\
 \rho_\beta(x)&=&\int dx \,\Omega(x)/2+\tilde C_\beta.
\end{eqnarray}
These solutions generalise the well-known space-dependent solutions $\rho_\text{LK}(x)$ as mentioned earlier. Instead of linear $x$-dependence, we now obtain general, nonlinear $x$-dependent solutions. As in the original LK-TASEP model, these solutions may meet each other or with the other solution $\rho=1/2$ in the bulk of the system giving phase coexistence of different phases in the steady states. Following the logic outlined in Ref.~\cite{frey-lktasep}, the steady state density profiles and ensuing phase diagram may be calculated by equating the steady state currents. It is easy to see that the resulting phase diagram generally will have the same topology as in the LK-TASEP model, although the precise locations of the phase boundaries in the $\alpha-\beta$ plane should depend on the specific forms of $\Omega(x)$. This reveals a degree of robustness of the phase diagrams in the LK-TASEP model, revealing universality in the topology of the phase diagrams.

\end{document}